\documentclass[9pt,twocolumn,twoside]{gsajnl}
\articletype{inv} 
\usepackage{lmodern}
\usepackage{amsmath,amssymb,amsfonts}
\newtheorem{definition}{Definition}
\usepackage{subcaption}
\usepackage{hyperref}
\usepackage{cleveref}

\runningtitle{Karim and Decker \textit{et al.}, Explainable AI for Bioinformatics} 
\runningauthor{Karim and Decker \textit{et al.}}

\title{Explainable AI for Bioinformatics: Methods, Tools, and Applications}

\author[1,2$\dagger$]{Md. Rezaul Karim}
\author[1]{Tanhim Islam}
\author[3, 1]{Oya Beyan}
\author[1,2]{Christoph Lange}
\author[4,5]{Michael Cochez}
\author[6,7]{Dietrich Rebholz-Schuhmann}
\author[1,2]{Stefan Decker}

\affil[1] {Department of Data Science and Artificial Intelligence, Fraunhofer FIT, Germany }
\affil[2] {Chair of Computer Science 5 - Information Systems and Databases, RWTH Aachen University, Germany }
\affil[3] {University of Cologne, Faculty of Medicine and University Hospital Cologne, Institute for Biomedical Informatics, Germany}
\affil[4] {Department of Computer Science, Vrije Univeriteit Amsterdam, the Netherlands}
\affil[5] {Elsevier Discovery Lab, Amsterdam, the Netherlands}
\affil[6] {ZBMED - Information Center for Life Sciences, Germany}
\affil[7] {Faculty of Medicine, University of Cologne, Germany}

\correspondingauthoraffiliation[$\ast$]{\textbf{Corresponding author}:~\url{rezaul.karim@rwth-aachen.de}. This paper is currently under review in Briefings in Bioinformatics.}

\begin{abstract}
    Artificial intelligence~(AI) systems utilizing deep neural networks~(DNNs) and machine learning~(ML) algorithms are widely used for solving important problems in bioinformatics, biomedical informatics, and precision medicine. However, complex DNNs or ML models, which are often perceived as opaque and \emph{black-box}, can make it difficult to understand the reasoning behind their decisions. This lack of transparency can be a challenge for both end-users and decision-makers, as well as AI developers. Additionally, in sensitive areas like healthcare, explainability and accountability are not only desirable but also legally required for AI systems that can have a significant impact on human lives. Fairness is another growing concern, as algorithmic decisions should not show bias or discrimination towards certain groups or individuals based on sensitive attributes. Explainable artificial intelligence~(XAI) aims to overcome the opaqueness of black-box models and provide transparency in how AI systems make decisions. Interpretable ML models can explain how they make predictions and the factors that influence their outcomes. However, most state-of-the-art interpretable ML methods are domain-agnostic and evolved from fields like computer vision, automated reasoning, or statistics, making direct application to bioinformatics problems challenging without customization and domain-specific adaptation. In this paper, we discuss the importance of explainability in the context of bioinformatics,  provide an overview of model-specific and model-agnostic interpretable ML methods and tools, and outline their potential caveats and drawbacks. Besides, we discuss how to customize existing interpretable ML methods for bioinformatics problems. Nevertheless, we demonstrate how XAI methods can improve transparency through case studies in bioimaging, cancer genomics, and text mining~(GitHub repo: \url{https://github.com/rezacsedu/XAI-for-bioinformatics}. Our review aims to provide valuable insights and serve as a starting point for researchers looking to enhance explainability and decision transparency in solving bioinformatics problems.
\end{abstract}


\dates{\rec{xx xx, xxxx} \acc{xx xx, xxxx}}

\begin{document}

\maketitle
\thispagestyle{firststyle}
\vspace{-13pt}

\section{Introduction}
Artificial intelligence~(AI) systems that are built on machine learning~(ML) and deep neural networks~(DNNs) are increasingly deployed in numerous application domains such as military, cybersecurity, healthcare, etc. Further, ML and DNN models are applied to solving complex and emerging biomedical research problems: from text mining, drug discovery, and single-cell RNA sequencing to early disease diagnosis and prognosis. 
Further, the paradigm of evidence-based precision medicine has evolved toward a more comprehensive analysis of disease phenotype and their induction through underlying molecular mechanisms and pathway regulation. 
Another common application of AI in precision medicine is predicting what treatment protocols are likely to succeed on a patient based on patient phenotype, demographic, and treatment contexts~\cite{davenport2019potential}. 

Biomedical data science includes various types of data like genome sequences, omics, imaging, clinical, and structured/unstructured biomedical texts~\cite{han2022challenges}, where ML methods are typically used for the analysis and interpretation of multimodal data~(e.g., multi-omics, imaging, clinical, medication, disease progression), in a multimodal data setting. Further, datasets, including bioimaging and omics are of increasing dimensionality. The surge of these massive amounts of data not only brings unprecedented progress in bioinformatics and opportunities to perform predictive modelling at scale~\cite{han2022challenges}, it also introduces challenges to existing AI methods and tools such as data heterogeneity, high-dimensionality, and volume~\cite{karim2020deep}.
Principal component analysis~(PCA) and isometric feature mapping~(Isomap) are extensively used as dimensionality reduction techniques~\cite{fournier2019empirical}. However, the representations learned by these methods often lose essential properties~\cite{clusteringAlgo}, making them less effective against a known phenomenon called \emph{curse of dimensionality}, particularly in high-dimensional datasets~\cite{fournier2019empirical}. 


A complex DNN model can effectively handle complex problems, thanks to its ability to extract features and to learn useful representations from high-dimensional datasets. For instance, autoencoders~(AEs) is used for unsupervised learning tasks, where their multi-layered, non-linear architecture enables the learning of complex and higher-order feature interactions. By transforming input feature space into a lower-dimensional latent space, AEs capture important contextual information of the underlying data, which can be used for various downstream tasks. However, the latent factors learned by AEs are not easily interpretable. disentangling them can provide insights into the features captured by the representations and the attributes of the samples that the tasks are based on~\cite{karim2020deep}.


AI has already surpassed human medical experts in specific areas such as detecting tumors and analyzing disease progression. However, the widespread use of AI in healthcare is hindered by the lack of models capable of handling vast amounts of data. Although DNN models can address complex problems, their \emph{black-box} nature raises concerns about transparency and accountability. With their increasing complexity, complex DNN models tend to be less and less interpretable and may end up as \emph{black-box} methods. Although mathematical certainty means it should be possible to transparently show no hidden logic is influencing the behaviour of a model~\cite{molnar2020interpretable}, predictions made by these models cannot be traced back, making it unclear how or why they arrived at a certain outcome. This lack of explainability can lead to issues of trust in the AI system and the inability to provide human users with explanations for its decisions.

\begin{figure}[h]
	\centering
		\includegraphics[width=0.45\textwidth]{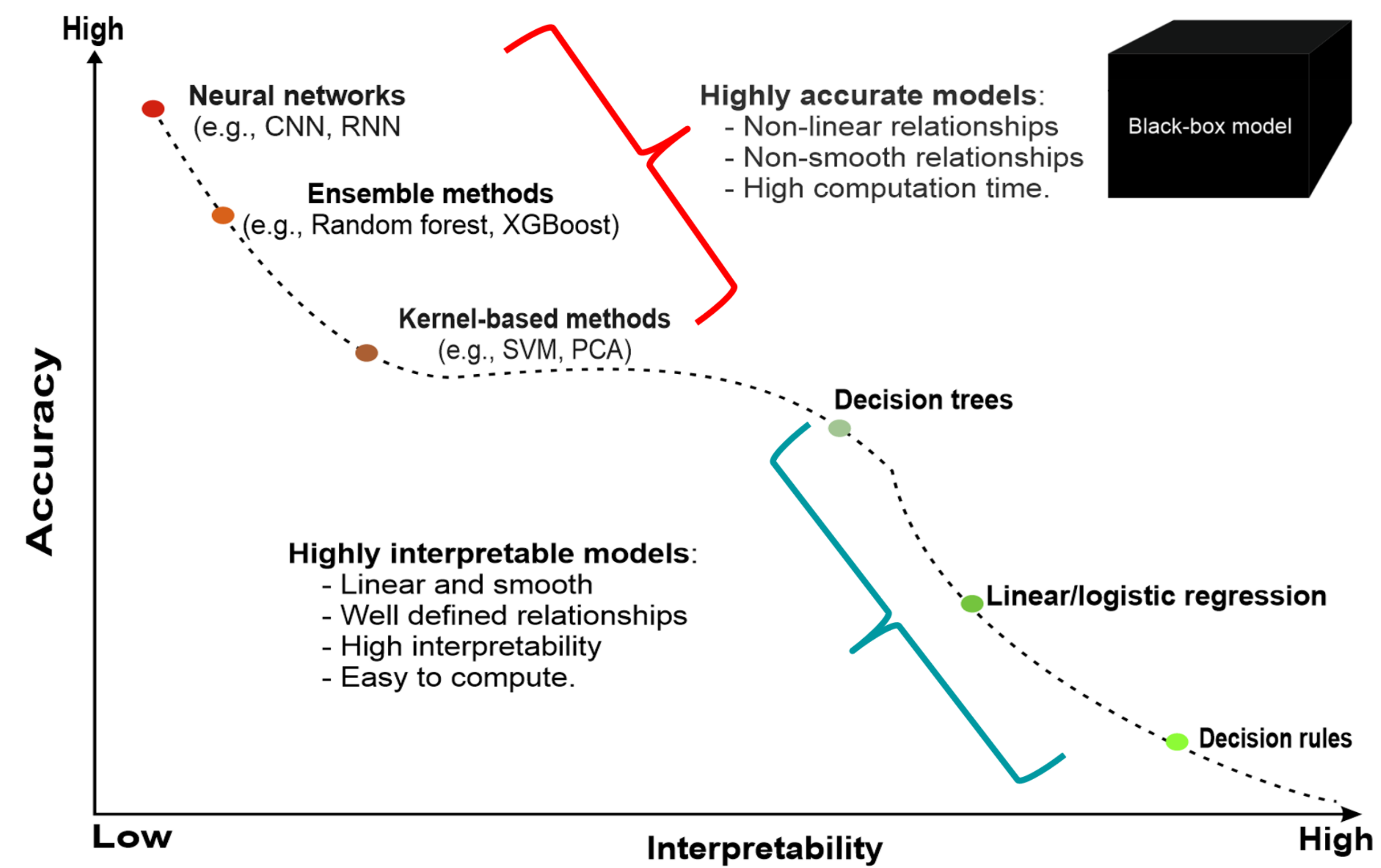}
		\caption{Accuracy vs. interpretability trade-off - complex models tend to be less and less interpretable~\cite{karim_phd_thesis_2022}}
        \label{fig:acc_vs_xai}
\end{figure}

The field of explainable artificial intelligence~(XAI) aims to make AI systems more transparent and understandable by interpreting how \emph{black-box} models make decisions. XAI strives to enhance the human-comprehensibility, reasoning, transparency, and accountability of AI systems~\cite{han2022challenges}. An interpretable machine learning~(ML) model can reveal the factors that impact~(e.g., statistically significant features) its outcomes and explain the interactions among these factors. While linear models, decision trees~(DTs), and rule-based systems are less complex and more interpretable, they are less accurate compared to tree-based ensembles and DNNs, which have been demonstrated to surpass simpler models. This phenomenon, which is shown in \cref{fig:acc_vs_xai} called the \emph{accuracy vs.\ interpretability trade-off}.  

The XAI has recently gained widespread interest from both academic and industry~\cite{tjoa2020survey}. This has led to the development of several model-specific and model-agnostic interpretable ML methods designed to enhance the local and global interpretability of ML models~\cite{wachter2017counterfactual}. Despite the advancements in the development of interpretable ML methods to explain the decisions of \emph{black-box} models in recent years, many of those methods are seldom used beyond visualization~\cite{weber2022beyond}. Besides, many of these methods are not designed in a domain-agnostic manner and may not be suitable for all biomedical or bioinformatics problems and data types~\cite{han2022challenges}. Therefore, they require customization or extension to fit the specific needs of bioinformatics and diverse data types~\cite{han2022challenges}. This paper outlines the importance and benefits of XAI, with a focus on bioinformatics. It provides an overview of various interpretable ML methods and shows, through several case studies in bioimaging, cancer genomics, and biomedical text mining, how bioinformatics research can benefit from XAI methods and improve decision fairness. 


\section{Importance of XAI in Bioinformatics}\label{sec:needs}
Handling large-scale biomedical data involves significant challenges, including heterogeneity, high dimensionality, unstructured data, and high levels of noise and uncertainty. Despite its data-driven nature and technical complexity, the adoption of data-driven approaches in many bioinformatics scenarios is hindered by the lack of efficient ML models capable of tackling these challenges. In order for AI systems to provide trustworthy and reliable decisions, the need for interpretable ML models has become increasingly important to ensure transparency, fairness, and accountability in critical situations~\cite{das2020opportunities}. 

\begin{figure*}
	\centering
	\includegraphics[width=0.75\textwidth]{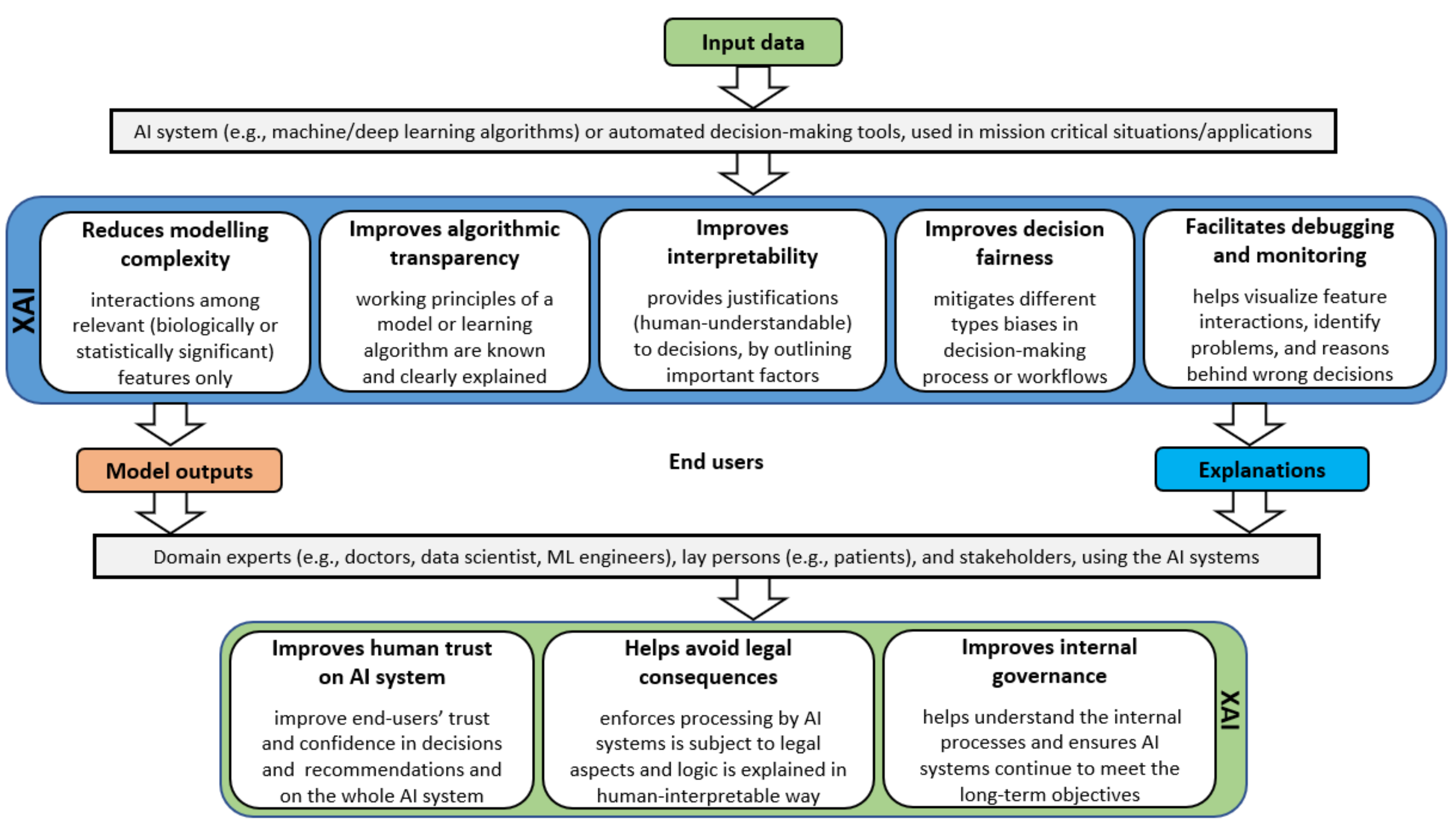}	
    \caption{Advantages of XAI in improving interpretability, trust, and transparency in algorithmic decision-making process}	
	\label{fig:xai_importance}
\end{figure*}

Although not all predictions need to be explained, having a model that is interpretable can make it easier for users to understand and trust its decisions~\cite{stiglic2020interpretability}. 
Weber \textit{et al.}~\cite{weber2022beyond} showed via some experiments that under the right conditions, augmentations based on XAI can provide significant, diverse, and reliable benefits advantages over \emph{black-box} models. We outline some key benefits of interpretable ML methods in \cref{fig:xai_importance}. 

\begin{figure*}[h]
	\centering
	\includegraphics[width=0.9\textwidth]{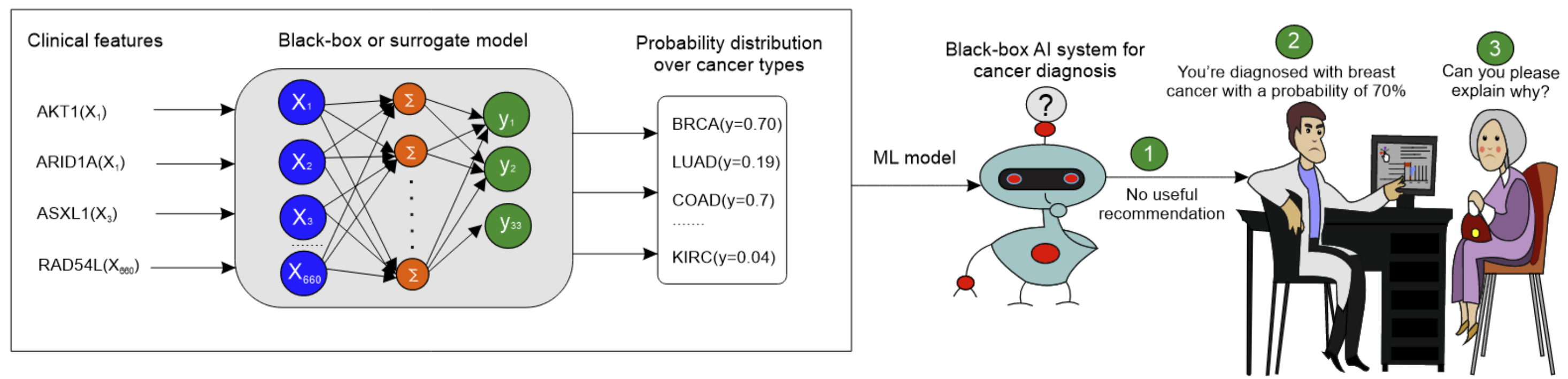}
	\caption{Example of practical consequence: a black-box model cannot explain diagnosis decisions~\cite{karim_phd_thesis_2022}}
    \label{fig:doctor_patient}
\end{figure*}

\subsection{Helps avoid practical consequences}

One of the critical applications of AI is aiding diagnosis and treatment of various cancerous conditions. Early detection and classification of patients into high or low-risk groups is crucial for effective management of the illness~\cite{kourou2015machine}. An example of this is a doctor diagnosing a patient with breast cancer. Given that breast cancer is a leading cause of death in women, it is important for the diagnosis to be thoroughly investigated. By utilizing omics data, such as genetic mutations, copy number variations~(CNVs), gene expression~(GE), DNA methylation, and miRNA expression, accurate diagnosis and treatment can be identified. Suppose a deep learning model trained on multi-omics data can classify cancerous samples from healthy samples with 95\% accuracy, and the model diagnoses a patient with a 70\% probability of having breast cancer. If the patient asks questions like \emph{``why do I have breast cancer?"} or \emph{``how did the model reach this decision?"} or \emph{``which biomarkers are responsible?"}, it may not be possible to clearly explain the model's decision-making process because the representations learned by the deep learning model may not be easily interpretable.

\begin{figure}
	\centering
	\includegraphics[width=0.5\textwidth]{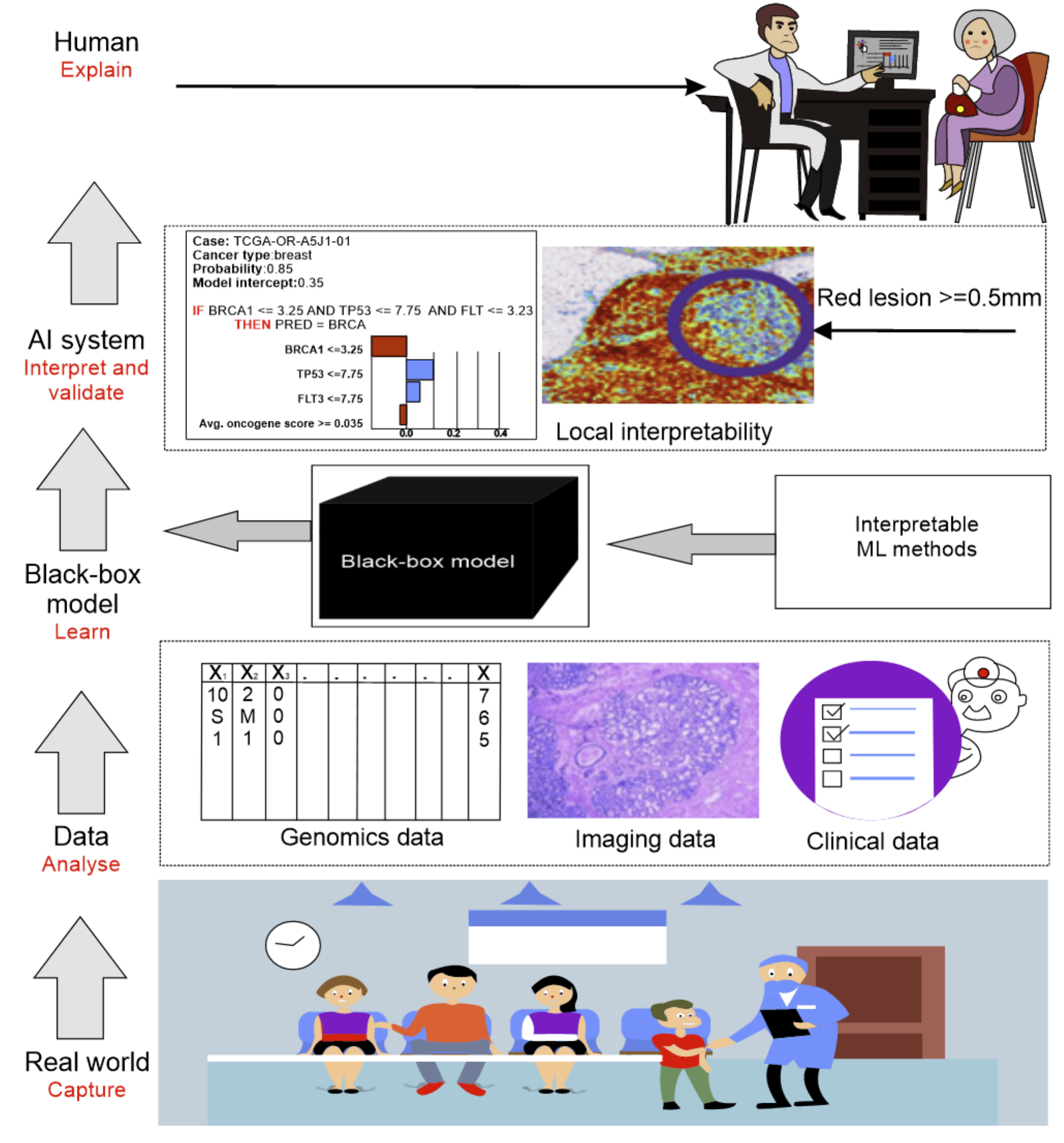}
	\caption{AI for cancer diagnosis in a clinical setting~\cite{karim_phd_thesis_2022}}
    \label{fig:real_Life_example}
\end{figure}

The diagnosis may further depend on several distinct molecular subtypes and factors like estrogen-, progesterone-, and human epidermal growth factor receptors. The diagnosis of a breast cancer patient requires a careful examination of multiple sources of data, including omics information, bioimaging, and clinical records. A multimodal DNN model trained on this data can classify samples with high accuracy. Further, image-guided pathology may need to employ in order to analyze imaging data~(e.g., histopathological image analysis), as shown in \cref{fig:real_Life_example}. However, the representations and decision-making process of such a multimodal model may not be easily interpreted. This can make it difficult to explain the diagnosis to the patient and raises concerns about the model's transparency and accountability in a clinical setting, as shown in \cref{fig:doctor_patient}. The use of black-box models is problematic as their internal logic is hidden from users, leading to theoretical, practical, and legal consequences~\cite{zednik2019solving}. Therefore, it is important for AI systems to have an explainable mode of operation.

An interpretable ML model, that emphasizes transparency and traceability of its logic, can explain why and how it arrived at certain decisions, reducing negative consequences. In the context of our cancer example, local interpretability can provide reasons for a decision made for a specific patient or reference to similar cases, allowing identification of unique characteristics of a patient in a small group~\cite{stiglic2020interpretability}. In contrast, global interpretability shows the overall behavior of the model at a high level, e.g., if an ML model is trained to predict gene up-regulation after treatment based on the presence of regulatory sequences, global interpretability will indicate the significance of the sequences in predicting up-regulation for all genes in the dataset, while local interpretability will reveal the importance of the sequences in predicting up-regulation for a specific gene.

\begin{figure*}[h]
	\centering
	\includegraphics[width=0.7\textwidth]{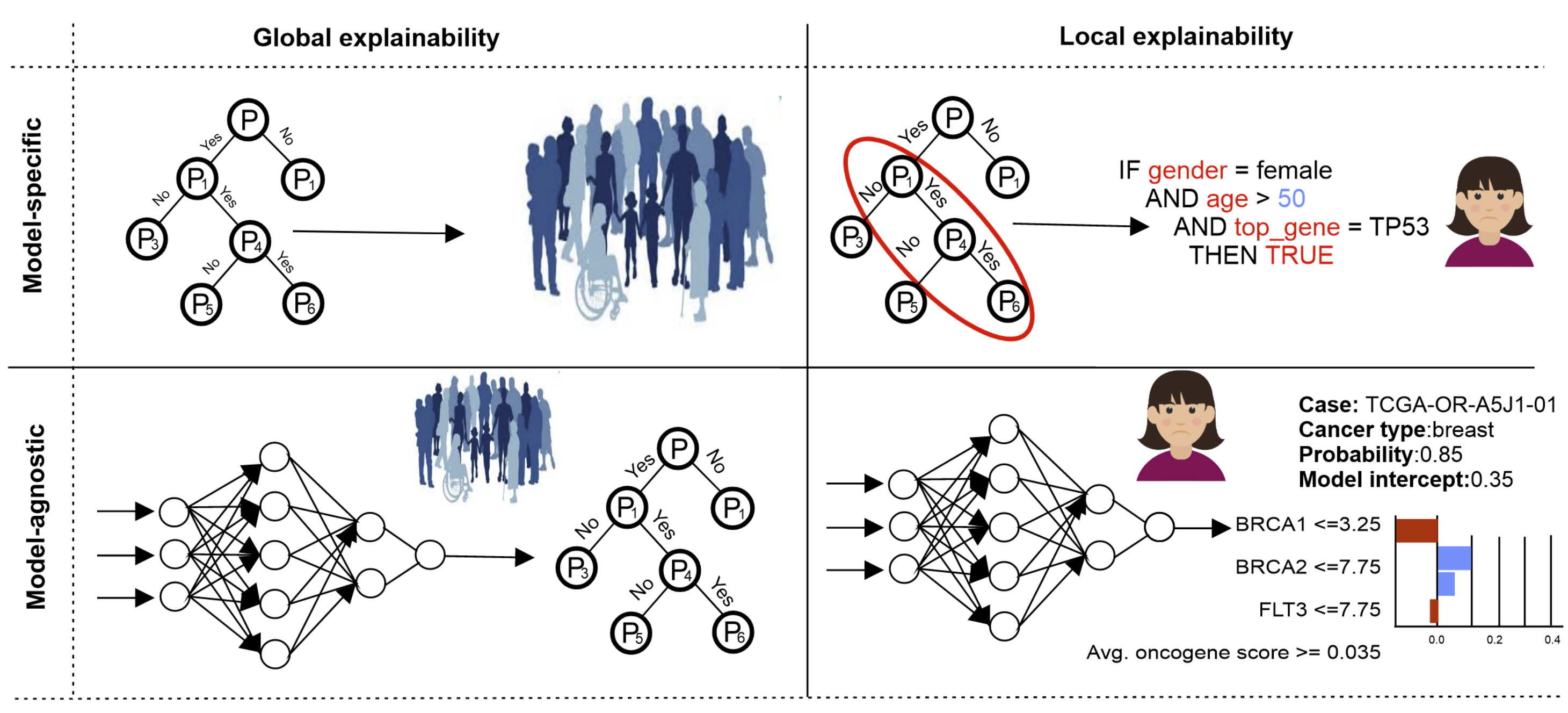}	
    \caption{An interpretable model can explain decisions locally, outlining the global behaviour~(conceptually based on~\cite{stiglic2020interpretability})}	
	\label{fig:local_vs_global_ex_2}
\end{figure*}

\subsection{Reduces complexity and improves accuracy}\label{subsec:biomarker_discovery}
The aim of genomics data analysis is to extract biologically relevant information and gain insight into the role of biomarkers, such as genes, in cancer development. However, biological processes are complex systems controlled by the interactions of thousands of genes, not single gene-based mechanisms~\cite{han2022challenges}. The high dimensionality of omics data, with many genes potentially irrelevant to the task of cancer prediction, creates challenges for ML models. With over 30,000 genes, the feature space for a model becomes very large, leading to sparse training data and small sample sizes in clinical trials. Including all features not only hinders the model's predictive power by adding unwanted noise, but also increases computational complexity.



Therefore, it is crucial to select biologically significant features with high correlation to the target classes and low correlation between genes. Accurately identifying cancer-specific biomarkers: i) enhances classification accuracy, ii) enables biologists to study the interactions of relevant genes, and iii) helps understand their functional behavior, leading to further gene discovery~\cite{karim2022interpreting}. After identifying these biomarkers based on feature attributions, they can be ranked based on their relative or absolute importance. These identified genes can serve as cancer-specific marker genes that distinguish specific or multiple tumor classes~\cite{karim2019onconetexplainer}. 

\subsection{Improves decision fairness}
With the widespread use of AI, it is essential to address fairness concerns, as AI systems can make significant and impactful decisions in sensitive environments~\cite{stiglic2020interpretability}. Bias is a major hindrance to fair decision-making and has been a subject of discussion in philosophy and psychology for a long time~\cite{fairness_survey}. Statistically, bias means a false representation of the truth with respect to the population~\cite{fairness_survey}, and it can occur at any stage of the ML pipeline, from data collection, feature selection, model training, hyperparameter setting, to interpretation of results for affected individuals, such as patients~\cite{GoogleBiasinMLPipeline}. For instance, in healthcare and biomedicine, representative, selection, and discriminatory biases can easily be present in biophysical data, raising fairness concerns and potentially leading to unfair outcomes in various learning tasks~\cite{fairness_survey}. ML algorithms are a type of statistical discrimination that may become problematic when they give certain privileged groups a systematic advantage and certain disadvantaged groups a systematic disadvantage. 

If the data used to train an ML model is biased, the model will also be biased and produce biased decisions if used in an AI system. A recent study~\cite{obermeyer2019dissecting} found that a widely used algorithm exhibits significant racial bias and affects millions of patients. The study showed that at a given risk score, black patients are significantly sicker than white patients and that improving this disparity would increase the number of black patients receiving additional help from 17.7\% to 46.5\%. Such scenarios can further erode trust in healthcare experts and other stakeholders. In the context of cancer, predictions based on a biased model can disproportionately impact the diagnosis, potentially leading to incorrect treatments.
As humans are inherently biased, data collected or prepared by humans will always contain some level of bias. It is important to be aware of common human biases that may appear in the data so that steps can be taken to reduce their impact before training an ML model. Making fair decisions requires a thorough understanding of the context, and interpretability can help identify factors that may lead to unfair outcomes~\cite{fairness_survey}. Once identified, proactive measures can be taken to ensure that the decisions don not discriminate against certain groups or populations because of those factors~\cite{mehrabi2019survey}.

\subsection{Internal governance and legal compliance}

As AI becomes more widespread, the need for transparency and explanation of AI decisions grows for ethical, legal, and safety reasons~\cite{das2020opportunities}. In sensitive domains like healthcare, where AI can impact human lives, explainability and accountability are not only desirable, but also legally required. The EU General Data Protection Regulation~(EU GDPR) recognizes the importance of ethics, accountability, and robustness in AI, and requires that automated decision-making processes have suitable safeguards, including the right to understand and challenge the decision, and the right not to be solely subject to an AI-based decision that significantly impacts one's life. Further, it enforces that processing based on automated decision-making tools should be subject to suitable safeguards, including \textit{``right to obtain an explanation of the decision reached after such assessment and to challenge the decision''} and individuals \emph{``have the right not to be subject to a decision based solely on automated processing and whenever human subjects have their lives significantly impacted by an automatic decision-making machine''}~\cite{kaminski2019right}. 


Given the central role AI plays in interactions between organizations and individuals~\cite{meske2022explainable}, it is important for AI-assisted decisions to be explained in a way understandable to those affected~\cite{kazim2020explaining}. The GDPR prohibits the use of AI for automated decision-making unless the logic involved is clearly explained. To avoid any legal consequences, the decision-making process should be made as transparent as possible through interpretability~\cite{das2020opportunities}. By making explainability a requirement, the organization will have greater understanding of what the AI system does and why, which will improve oversight and increase precision. This also helps the organization comply with parts of GDPR and adhere to external policies, practical consequences, and processes that regulate business practices~\cite {kazim2020explaining}.

\section{Techniques and Methods for Interpretable ML} \label{sec:xai_methods}
A wide range of model-specific and model-agnostic interpretable ML methods have been proposed and developed~\cite{azodi2020opening}. All these methods largely fall into three main categories: probing, perturbing, and model surrogation. Further, depending on the level of abstractions, they can be categorized as local interpretability and global interpretability methods. We provide a list of papers, methods, and tools as well as supporting Jupyter notebooks, covering bioimaging, cancer genomics, text mining, and reasoning examples. We categorize these interpretable ML methods based on papers, books, and tools. 

\subsection{Terminologies and notations}
Interpretability and explainability are interchangeably used in literature. However, the former signifies cause and effect of an outcome in a system, the latter is the extent to which the internal working mechanism of an AI system can be explained. According to Cambridge Dictionary\footnote{\scriptsize{\url{https://dictionary.cambridge.org/dictionary/english/interpretable}}}, \emph{``if something is interpretable, it is possible to find its meaning or possible to find particular meaning in it''}. Miller \textit{et al.}~\cite{miller2018explanation} define \emph{explanation} as the answer to `why' questions. Das \textit{et al.}~\cite{das2020opportunities} define interpretation as a \emph{``simplified representation of a complex domain, such as outputs generated by an ML model, to meaningful, human-understandable, and reasonable concepts''}. Interpretability \emph{``is the degree to which a human can understand the cause of a decision''}~\cite{miller2018explanation}. 
Interpretability of the ML model is the extent to which the cause and effect can be observed. 

\begin{table*}
    \centering
    \caption{Interpretable ML techniques, categorized based on agnosticism, scopes, underlying log, and data types}
    \scriptsize{
    \begin{tabular}{l|l|l|l|l}
    \hline
        \textbf{Approach} & \textbf{Scope} & \textbf{Agnosticism} & \textbf{Methodology} & \textbf{Datatype} \\ \hline
        CAM~\cite{he2016computer, izadyyazdanabadi2018weakly} & Local & Model agnostic & Back-propagation & Image \\ \hline
        Grad-CAM~\cite{selvaraju2017grad} & Local &  Model agnostic & Back-propagation & Image \\ \hline
        Grad-CAM++~\cite{chattopadhay2018grad} & Local & Model agnostic & Back-propagation & Image \\ \hline
        Guided Grad-CAM~\cite{tang2019interpretable} & Local & Model agnostic & Back-propagation & Image \\ \hline
        Respond CAM~\cite{zhao2018respond} & Local &  Model agnostic & Back-propagation & Image \\ \hline
        LRP~\cite{LRP1,bach2015pixel} & Local &  Model agnostic & Back-propagation & Image \\ \hline
        DeepLIFT~\cite{shrikumar2017learning} & Global/local &  Model agnostic & Back-propagation & Image \\ \hline
        Prediction difference analysis~(PDA)~\cite{zintgraf2017visualizing} & Local & Model agnostic & Perturbation & Image \\ \hline
        Slot activation vectors~\cite{jacovi2018understanding} & Global & Model agnostic & Back-propagation & Text \\ \hline
        Peak response mapping~(PRM)~\cite{zhou2018weakly} & Local & Model agnostic & Back-propagation & Image \\ \hline
        LIME~\cite{LIME,ribeiro2016should} & Global/local & Model agnostic & Perturbation & Image, text, tabular \\ \hline
        MUSE~\cite{lakkaraju2019faithful} & Global & Model agnostic & Perturbation & Text \\ \hline
        DeConvolutional nets~\cite{zeiler2014visualizing} & Local & Model agnostic & Back-propagation & Image \\ \hline
        Guided back-propagation~\cite{springenberg2014striving, izadyyazdanabadi2018weakly} & Local & Model agnostic & Back-propagation & Image \\ \hline
        Activation maximization~\cite{erhan2010understanding} & Local & Model agnostic & Back-propagation & Image \\ \hline
        Bayesian averaging over decision trees~\cite{schetinin2007confident} & Global & Model specific & Bayesian & Tabular \\ \hline
        Generative discriminative models~(GDM)~\cite{caruana2015intelligible} & Global & Model specific & Discriminative & Tabular \\ \hline
        Bayes rule lists~\cite{letham2015interpretable} & Global & Model specific & Rule-based & Tabular \\ \hline
        Shapley sampling~\cite{SHAP} & Global/local & Model agnostic & Perturbation & Image, text, tabular \\ \hline
        Gradient-based saliency maps~\cite{simonyan2013deep} & Local & Model agnostic & Back-propagation & Image \\ \hline
        Bayesian case model~(BCM)~\cite{kim2014bayesian} & Global & Model specific & Bayesian & Image, text, tabular \\ \hline
        Deep Taylor decomposition~\cite{montavon2017explaining} & Local & Model agnostic & Decomposition & Image \\ \hline
        Deep attribution maps~\cite{ancona2017towards} & Local & Model agnostic & Back-propagation & Image, text \\ \hline
        Axiomatic attributions~\cite{sundararajan2017axiomatic} & Local & Model agnostic & Back-propagation & Image, text \\ \hline
        PatternNet and pattern attribution~\cite{kindermans2017learning} & Local & Model agnostic & Back-propagation & Image \\ \hline 
        Neural additive models~(NAMs)~\cite{agarwal2021neural} & Global & Model specific & Others & Image \\ \hline
        ProtoAttend~\cite{arik2019protoattend} & Global & Model agnostic & Others & Image \\ \hline
        Concept activation vectors~(CAVs)~\cite{kim2018interpretability} & Global & Model agnostic & Others & Image \\ \hline
        Global attribution mapping~(GAM)~\cite{ibrahim2019global} & Global & Model agnostic & Perturbation & Image \\ \hline
        Randomized sampling for explanation~(RISE)~\cite{petsiuk2018rise} & Local & Model agnostic & Perturbation & Image \\ \hline
        Spectral relevance analysis~(SpRAy)~\cite{lapuschkin2019unmasking} & Global & Model agnostic & Back-propagation & Image \\ \hline
        Salient relevance~(SR) map~\cite{li2019beyond, wang2021saled} & Global/local & Model agnostic & Back-propagation & Image \\ \hline
        Automatic concept-based explanations~(ACE)~\cite{ghorbani2019towards} & Global & Model agnostic & Others & Image \\ \hline
        Causal concept effect(CaCE)~\cite{goyal2019explaining} & Global & Model agnostic & Others & Image \\ \hline
    \end{tabular}}
    \label{tab:methods_scope_agnosticism_methodology_hi_dt_xai}
\end{table*}

An interpretable ML refers to methods that make the behaviour and predictions of a system understandable to humans~\cite{molnar2020interpretable}. Algorithmic transparency suggests \emph{factors that influence the decisions made by algorithms should be visible, or transparent, to the people who use, regulate, and are affected by systems that employ those algorithms}~\cite{diakopoulos2017algorithmic}. 
An interpretable model can outline how input instances are mapped to certain outputs by identifying statistically significant features. 
While explainability is using the knowledge of what those features represented and their relative importance in explaining the predictions to humans in an understandable way or terms. Further, we refer to following terminologies and notations from our early work~\cite{karim2022interpreting} to understand several concepts used in this paper.

\begin{definition}[\textbf{Dataset}] \label{def:dataset}
    $D=(\tilde{X}, \tilde {Y})$ is a dataset, where $\tilde{X}$ be an $N$-tuple of $M$-instances, ${X}$ be the set of all instances in $\tilde{X}$, and $\tilde{Y}$ be $N$-tuple of labels $l \in L$. 
\end{definition}

\begin{definition}[\textbf{Model and prediction}] \label{def:classifier_and_prediction}
    Let the pair $(name, value)$ be a parameter and $\Theta$ be a set of parameters. A \emph{model} $f$ is a parametric function $f:X \times \Theta \rightarrow {\mathbb{R}}$ that maps an input instance $x$ from its feature space $X$ to a decision $y \in L$ and returns a real-valued output called \emph{prediction} $\hat y$. A prediction $\hat y=f(x_i, \theta)$ is accurate for model $f$ and parameter $\theta \in \Theta$ if and only if $\hat y = \tilde{Y}\left[i\right]$~(classification) or $\hat y \approx \tilde{Y}\left[i\right]$~(regression) for $x=\tilde{X}\left[i\right]$, where $1 \leq i \leq M$. 
\end{definition}

\begin{definition}[\textbf{Black-box and interpretable models}] \label{def:black_box_and_white_box_models}
    Let $f$ be a model and $\Theta$ be a set of parameters. Model, $f$ is a {black-box} if its internal working principle and ${\theta}$ are hidden or uninterpretable by humans owing to lack of traceability of how $f$ makes predictions. Model $f$ is interpretable if the parameters ${\theta}$ are known and there exists a mathematical interpretation $\lambda_i$ showing how and why a certain prediction $\hat y$ is generated by $f$.  
\end{definition}

\begin{definition}[\textbf{Algorithmic transparency}] \label{def:algo_trans_1} 
    A model $f$ is transparent if there exists a mathematical interpretation $\lambda_t$ by which a learning algorithm learns model $f$ by mapping relations between $X$ and $Y$ to make predictions.
\end{definition}

\begin{definition}[\textbf{Feature importance}] \label{def:feature_importance}
    Let $a_i$ be a feature in instance $x$ and $A$ be the set of all features. Importance function $h: A \rightarrow [0,1]$ assigns each element of $A$ a non-negative number between 0 and 1: the larger the number, the higher the importance of $a_i$. Local feature importance for $x$ is a set of feature and importance pairs $I_x=\left \{(a_1, h(a_1)), (a_2, h(a_2)), \dots, (a_M, h(a_M))\right\}$ for all $a_i \in x$. {Global feature importance} for $X$ is a set of feature and importance pairs $\bar I_X=\left \{(a_1, \bar p_1), (a_2, \bar p_2), \dots, (a_k, \bar p_M)\right\}$, where $\bar p_i$ is the mean local feature importance of $a_i$. 
\end{definition}

\begin{definition}[\textbf{Feature impacts}] \label{def:feature_impacts}
    Let $a_i$ be a feature of instance $x$, and $A$ is the set of all feature names. Impact function $g: A \rightarrow [-1, 1]$ takes an element of $A$ as input and results in a real number between -1 and 1. Local feature impact for $x$ is a set of feature and impact pairs $T_x=\left \{(a_1, g(a_1)), (a_2, g(a_2)), \dots, (a_M, g(a_M))\right\}$ for all $a_i \in x$. Global feature impact for $X$ is a set of feature and impact pairs $\bar T_X=\left \{(a_1, \bar q_1), (a_2, \bar q_2), \dots, (a_k, \bar q_M)\right\}$, where $\bar q_i$ is the mean of all local feature impacts of feature for $a_i$. 
\end{definition}

\begin{definition}[\textbf{Top-k and bottom-k features}] \label{def:top_k_important_features}
    Let $f$ be a model, $I$ be {global feature importance} for $D$, and $k$ be an integer. The {top-$k$ features} is a $k$-tuple such that for all $i \leq k \leq m$, $I[k[i]] \geq I[k[m+1]]$, and $I[k[i]] \geq I[k[i+1]]$, where $k$ is the number of top features used to explain the decision. A {bottom-k features} is a $k$-tuple such that for all $i \geq k \geq m$, $I[k[i]] \leq I[k[m+1]]$, and $I[k[i]] \leq I[k[i+1]]$\footnote{\scriptsize{Since the learning algorithm for a model involves stochasticity and the way they are computed could be different, feature importance scores~($p$) could be different if $I$ is sorted in ascending or descending orders of $p$.}}. 
\end{definition}

\begin{definition}[\textbf{Top-k and bottom-k impactful features}] \label{def:top_k_impactful_features}
    Let $f$ be a model, $T$ be {global feature impacts} for $D$, and $k$ be an integer. The {top-$k$ features} is a $k$-tuple such that for all $i \leq k \leq m$, $T[k[i]] \geq T[k[m+1]]$, and $T[k[i]] \geq T[k[i+1]]$, where $k$ is the number of top features used to explain a decision. A {bottom-k impactful features} is a $k$-tuple such that for all $i \geq k \geq m$, $T[k[i]] \leq T[k[m+1]]$, and $T[k[i]] \leq T[k[i+1]]$\footnote{\scriptsize{Since the learning algorithm for a model involves stochasticity and the way they are computed could be different, feature impact scores~($q$) could be different if $T$ is sorted in ascending or descending orders of $q$.}}.  
\end{definition}

\begin{definition}[\textbf{Interpretability}] \label{def:interpretability}
    \emph{Interpretability} is the degree to which humans can understand the cause of a decision~\cite{miller2018explanation}.  
    Global interpretability refers to an explanation $E$ explaining why model $f$ has predicted $\hat Y$ for all instances in $X$, outlining conditional interactions between dependent- and independent variables using some functions $\sigma(\cdot,\cdot)$. Local interpretability $e$ reasons why $\hat y$ has been predicted for an instance $x$, showing conditional interactions between dependent variables and $\hat y$, focusing on individual predictions. 
\end{definition}

\begin{definition}[\textbf{Algorithmic fairness}] \label{def:algo_fairness} 
    An algorithm is fair if the predictions do not favour or discriminate against certain individuals or groups based on sensitive attributes. 
\end{definition}

\begin{definition}[\textbf{Explanations}] \label{def:explanation}
    An explanation $e$ for a prediction $\hat y = f(x)$ for an instance $x \in X$ is an object derived from model $f$ using some function $\sigma(\cdot,\cdot)$ that reasons over $f$ and $x$ for $y$ such that $e = \sigma(f,x)$ and $e \in E$, where $E$ is the human-interpretable domain and $\sigma$ is an explanation function. 
\end{definition}

\begin{definition}[\textbf{Decision rules}] \label{def:decision_rules}
    A {decision rule} $r$ is a formula $p_1 \wedge p_2 \wedge \cdots \wedge p_k \rightarrow y$, where $p_i$ are boolean conditions on feature values in an instance $x$ and $\hat y$ is the \emph{}{decision}. Each decision rule $r$ is evaluated w.r.t $x$ by replacing the feature with a feature value from $x$, evaluating the boolean condition and reaching a conclusion $y$ if the condition is evaluated to be true.  
\end{definition}

\begin{definition}[\textbf{W-perturbations}] \label{def:w_perturbation}
    Let $D=(\tilde{X}, \tilde {Y})$ be a dataset, let $X$ be the set of all instances in $\tilde{X}$, and let $x$ be an instance of an $m$-tuple in $X$. Let $x^\prime$ for $x$ be a resulting vector by applying minimum change $\Delta x$ to some feature values $v_i$ using an optimization method~(e.g., ADAM~\cite{wachter2017counterfactual}) such that $y^{\prime}=f(x^\prime)$ against original prediction $y=f({x})$, where $x^\prime = x + \Delta x$. Then, $x^\prime$ is called a ${W}$-perturbation of $x$ and $y^\prime$.  
\end{definition}

\begin{definition}[\textbf{Counterfactual rules}] \label{def:Counterfactual_rules}
    Let $r: p \rightarrow y$ be a decision rule for an instance $x$ and $x^{\prime}$ is the perturbed vector of $x$~(e.g., w-perturbation\footnote{\scriptsize{Making changes to certain features may lead to a different outcome.}}). A \textit{counterfactual rule} $r^\updownarrow$ for the boolean conditions $p$ is a rule of the form $r^\updownarrow: p[\delta] \rightarrow y^{\prime}$ for $y^{\prime}=f(x^{\prime})$ s.t. $y^{\prime} \neq y$, where $\delta$ is a \textit{counterfactual} for original decision $y=f(x)$ for model $f$. 
\end{definition}

\begin{definition}[\textbf{Surrogate model}] \label{def:surrogation}
    Let $x \in X$ be an instance of an $m$-tuple, $\tilde{Y}$ be $N$-tuple of labels $l \in L$, and $f_b$ be a black-box model. Model $f$ is a \emph{surrogate model} of $f_b$ if the variance $R^2$ of $f_b$ captured by $f$ is $\approx 1$. 
\end{definition}

\begin{definition}[\textbf{Causally interpretable model}] \label{def:casually_interpretable_model}
    Let $f$ be an interpretable model and $q$ be a causal or counterfactual question. Then, the model $f$ is casually interpretable if there exists at least one answer $\hat y_a$ to question $q$. 
\end{definition}

\begin{definition}[\textbf{Local explanations}] \label{def:local_explanation_and_domain}
     Let $f$ be an interpretable model, $\hat y=f(x)$ be the prediction for instance $x$, $r$ be a \emph{decision rule} for $\hat y$, $\Phi$ be the set of \emph{counterfactual rules} for $r$, and $I$ be the set of \emph{local feature importance} for $\hat y$. A local explanation $e$ explaining decision $\hat y$ for $x$ is a triple $(I, r,\Phi)$. The domain $E_\ell$ of $e$ is a $N$-tuple of triples $(e_1, e_2,\cdots, e_N)$, where $N$ is the number of instances in $X$. 
\end{definition}

\begin{definition}[\textbf{Global explanations}] \label{def:global_explanation_and_domain}
    Let $f$ be a model, $\hat y=f(x)$ be the prediction for instance $x$, $\bar I$ and $\bar T$ be the sets of global feature \emph{importances} and \emph{impacts} for the set of predictions $\hat Y$ for instances $X$, $\bar I_{k_{t}} = \{(a_1, \bar q_{1_{t}}), (a_2, \bar q_{2_{t}}), \dots, (a_k, \bar q_{k_{t}})\}$ be the set of top-$k$ features, and $\bar I_{k_{b}} = \{(a_1, \bar q_{1_{b}}), (a_2, \bar q_{2_{b}}), \dots, (a_k, \bar q_{k_{b}})\}$ be the set of bottom-$k$ features. A global explanation $e_g$ is a pair $\langle \bar I_{k_{t}}, \bar I_{k_{b}} \rangle$. The domain $E_g$ of $e_g$ is a pair $(\bar I, \bar T)$. 
\end{definition}

The methods discussed in the following sections largely operate by approximating the outputs of a \emph{black-box} model via tractable logic or by approximating via a linear model. Besides, tree-, rules-, and knowledge-based interpretable methods have been proposed. 

\begin{figure*}
	\centering
	\begin{subfigure}{.48\linewidth}
		\centering
		\includegraphics[width=0.9\textwidth]{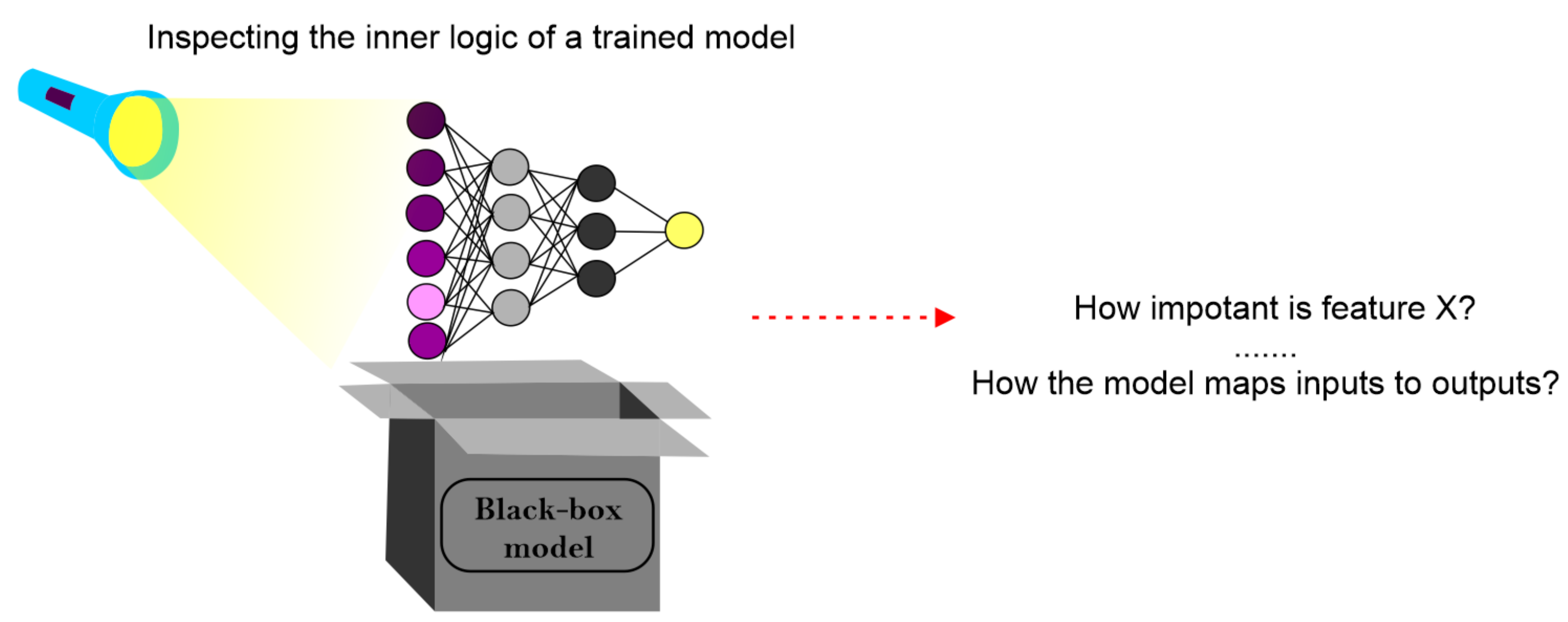}
		\caption{Probing a black-box model}
        \label{fig:probing}
	\end{subfigure}
	\hspace{2mm}
	\begin{subfigure}{0.48\linewidth}
		\centering
		\includegraphics[width=0.9\textwidth]{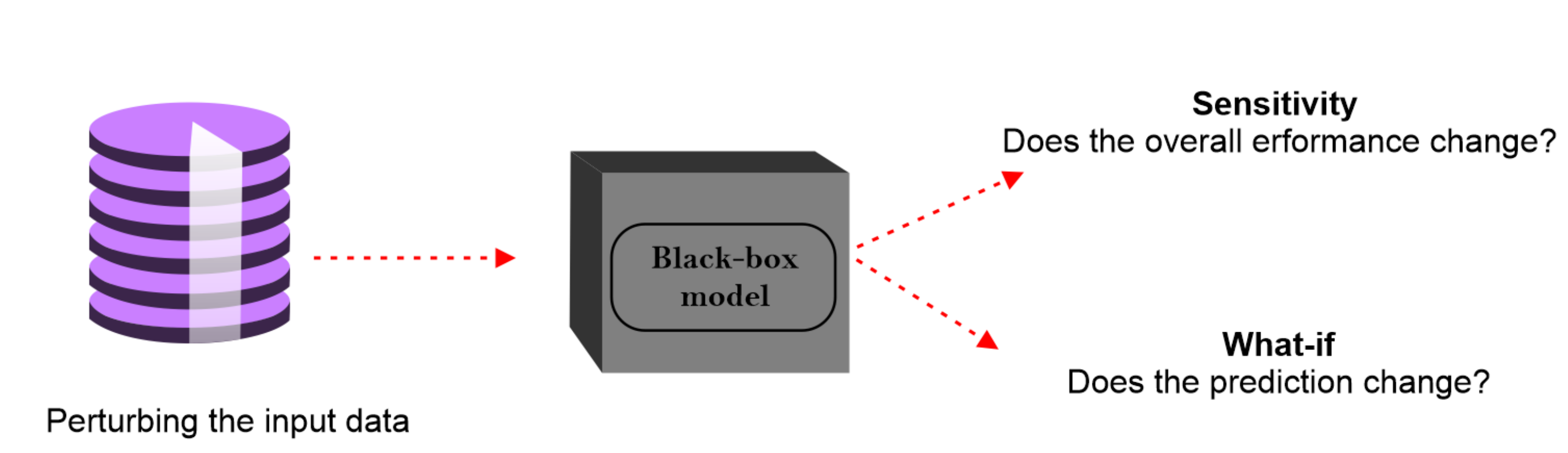}
		\caption{Interpreting black-box model with perturbing}
        \label{fig:purturbing}
	\end{subfigure}
		\begin{subfigure}{0.48\linewidth}
		\centering
		\includegraphics[width=0.9\textwidth]{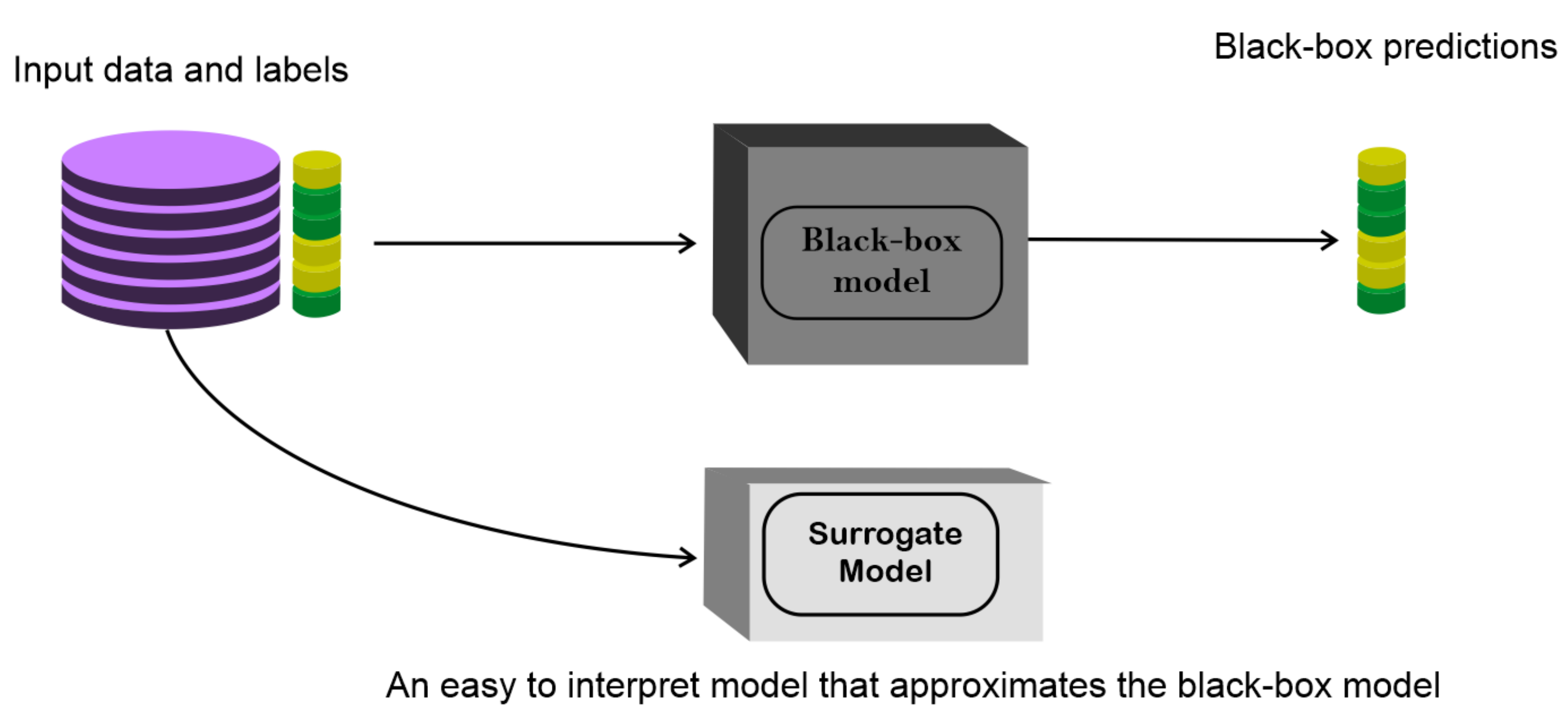}
		\caption{Interpreting black-box model with model surrogation}
        \label{fig:surrogation}
	\end{subfigure}
	\caption{Overview of probing, perturbing, and model surrogation interpretability methods~(based on Azodi \textit{et al.}~\cite{azodi2020opening})}
	\label{fig:pro_per_surroga}
\end{figure*}

\subsection{Probing black-box models} 
Some interpretable methods take into account the inner working principles of \emph{black-box} models. Therefore, probing, different probing techniques have been developed to understand their logic. Examples of probing techniques include gradient-based methods like gradient-weighted class activation mapping~(Grad-CAM++)~\cite{chattopadhay2018grad} and layer-wise relevance propagation~(LRP)~\cite{LRP1}, which use first-order gradient information from the black-box model to generate heatmaps indicating the relative importance of input features. These techniques are useful in bioimaging, where a convolutional neural network~(CNN) can learn features, e.g., class-discriminating pixels using filters and edge detectors across convolutional layers, and generate attention maps to highlight the most important pixels in an image~(e.g., pixels in a CT/MRI/X-Ray image).

\subsubsection{Saliency and gradient-based methods} 
 ~- saliency maps~(SMs) and gradient-based techniques are applied to locate crucial areas and then assign significance to each feature, for instance, a pixel in an image. These techniques include guided backpropagation~\cite{springenberg2014striving}, class activation maps~(CAM)~\cite{zhou2016learning}, Grad-CAM\cite{selvaraju2017grad}, Grad-CAM++~\cite{chattopadhay2018grad}, and LRP. In gradient-based methods, heatmaps~(HMs) are displayed using absolute output gradients and input nodes' negative gradients are zeroed out at the rectified linear layers of the network during the backward pass. This rectification of gradients results in more precise HMs~\cite{bohle2019layer}. The Class-discriminatory Attention Map is utilized to show the weighted combination of feature maps~(FMs). To highlight where a CNN focuses more, CAM computes weights for each FM based on the final convolutional layer. However, if the classifier is replaced with linear layers, the network must be retrained and the classifier's non-linearity disappears. 

\begin{figure*}
	\centering
	\includegraphics[width=\textwidth]{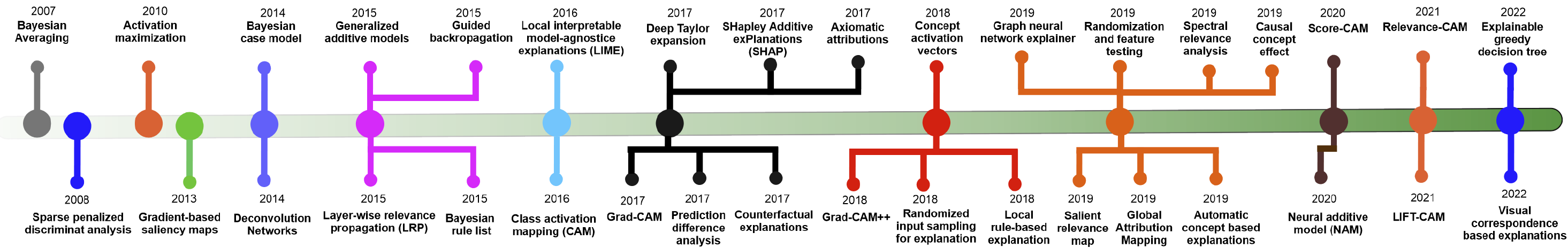}	
    \caption[Timeline of XAI methods and algorithms]{Timeline and evolution of interpretable ML methods, covering scopes, methodology, and usage level~\cite{karim_phd_thesis_2022}}	
	\label{fig:xai_timeline}
\end{figure*}


An improved version of CAM, called Grad-CAM~\cite{114}, has been proposed that uses globally average gradients of FMs as weights for target class $c$ instead of pooling. The guided backpropagation in Grad-CAM produces more interpretable but less class-sensitive visualizations compared to SMs. Since SMs use true gradients, network weights tend to impose a stronger bias towards specific input pixels. Grad-CAM highlights class-relevant pixels instead of generating random noise~\cite{nie2018theoretical} by using HMs to focus attention on and locate class-discriminating regions of an image. The class-specific weights for each FM are gathered from the final convolutional layer through globally averaged gradients~(GAG) instead of pooling~\cite{chattopadhay2018grad}.

\begin{figure*}
    \centering
    \includegraphics[width=0.8\textwidth]{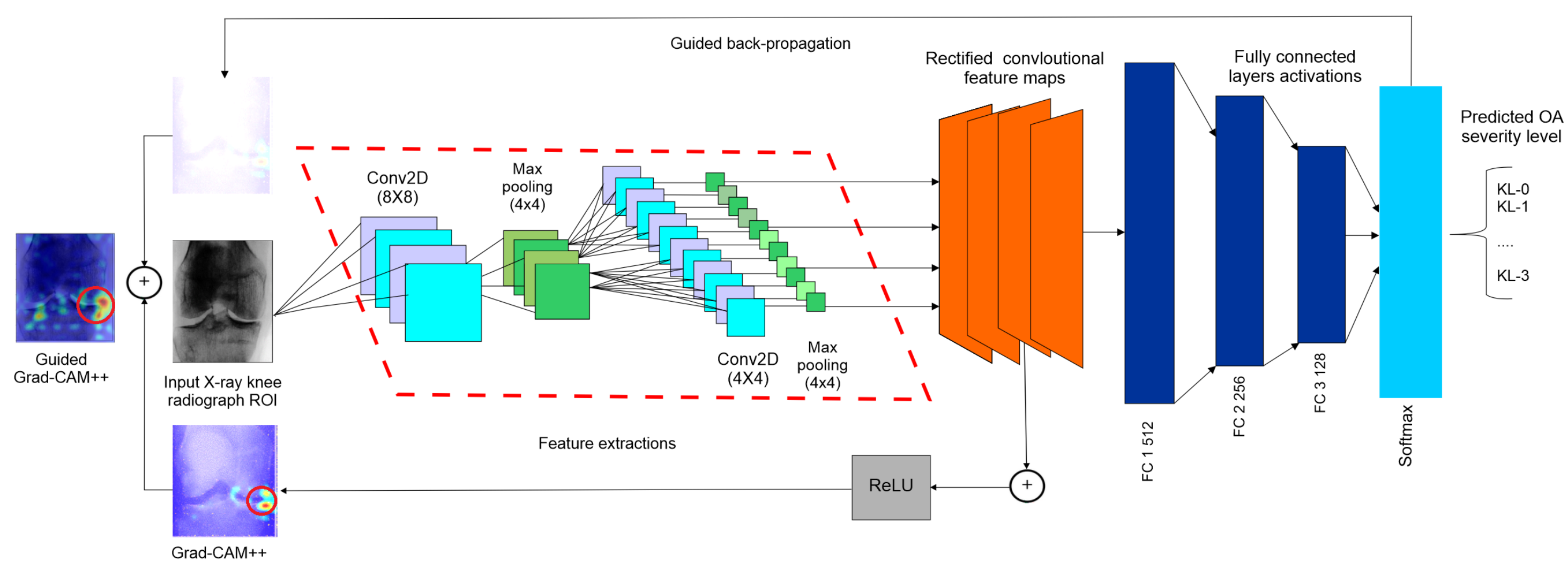}    
    \caption{Example of Grad-CAM++ showing pixel relevances to localize critical regions in knee radiography~(based on ~\cite{karim2021deepkneeexplainer})}    
    \label{fig:viz}
\end{figure*}

Grad-CAM has a limitation in visualizing multiple occurrences of the same class in an image with slightly different orientations, causing some objects to disappear from the SMs. This is due to its inability to recognize significance disparities among pixels, leading to parts of objects being rarely localized. To address this, Grad-CAM++ was introduced, which replaces the GAG with a weighted average of pixel-wise gradients. A typical example of using Grad-CAM++, as shown in \cref{fig:viz}, involves taking a radiograph as input, passing it through convolutional layers, rectifying the convolutional feature maps using guided backpropagation and Grad-CAM++, and then feeding it into a fully-connected softmax layer for classification. The critical areas of the image are localized using HMs. Although CAM variants back-propagate gradients to the inputs, gradients are actually propagated only up to the final convolutional layer, and they are specific to the architecture.

LRP is a technique suitable for imaging tasks, as it is based on the idea that the likelihood of a class can be traced backwards through a network to the individual layer-wise nodes of the input~\cite{LRP2}. For image recognition, LRP produces HMs that highlights the important pixels of an image for the model's prediction. This is achieved by running a backward pass in a CNN, which is a conservative redistribution of relevance, where nodes that contribute the most to the higher layers receive the most relevance. First, the image is classified through a forward pass, and then the relevance is back-propagated to generate a relevance map. The relevance of each node in a layer is calculated recursively, and if the node-level relevance is negative, it is calculated using the ReLU function~\cite{LRP2}.

\begin{figure*}
	\centering
	\includegraphics[width=0.8\textwidth]{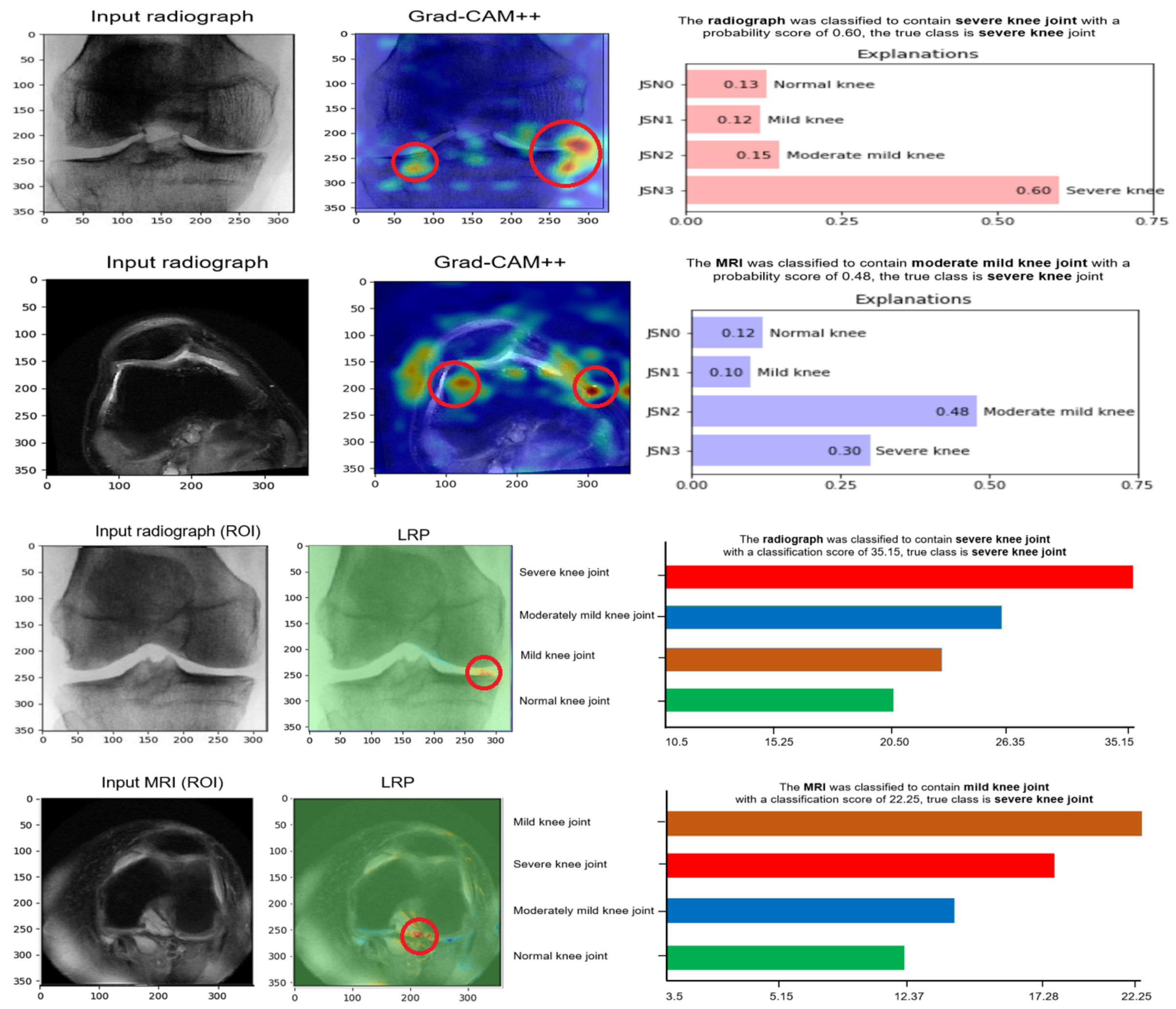}	
    \caption{Example of using Grad-CAM++ and LRP for osteoarthritis diagnosis from knee MRI and X-ray images, \\highlighting critical knee regions to emphasise plus textual explanations~(based on~\cite{karim2021deepkneeexplainer})}	
	\label{fig:explain_with_lrp}
\end{figure*}

While the regions highlighted by Grad-CAM++ for class discrimination are less precisely localized and scattered, LRP highlights them more accurately~\cite{karim2021deepkneeexplainer}. This is because Grad-CAM++ replaces the globally average gradient with a weighted mean, which highlights conjoined features more precisely. As a result, if a diagnosis has to be based on microscopic histopathology images, an AI-assisted image analysis tool could help confirm the presence of breast cancer. The class-discriminative regions can be localized with respect to pixel relevance, making Grad-CAM++ and LRP useful in improving the reliability of diagnoses. For instance, \cref{fig:explain_with_lrp} shows an example of an explainable diagnosis of osteoarthritis using MRT and X-ray images. As seen, both Grad-CAM++ and LRP generate reliable heat maps that highlight critical image regions and provide precise localization while marking similar regions as important.

\subsubsection{Attention-based probing techniques} 
~- attention mechanisms are designed to identify significant portions of features, leading to enhanced precision in various language modeling assignments. Propositional self-attention networks~(PSAN)~\cite{vaswani2017attention} is an early approach relying on attention, where attention heads symbolize different connections between input features. Transformer language models~(TLMs) such as bidirectional encoder representations from transformers~(BERT)~\cite{devlin2018bert}, employ attention to identify significant tokens for next word prediction by representing diverse connections between input features through bidirectional attention. BERT uses vast amounts of unsupervised data to produce context-sensitive representations~\cite{xue2019fine}. The attention technique is widely used in NLP, computer vision, and speech recognition. Research by Xu \textit{et al.}~\cite{xu2020building} has demonstrated the potential of TLMs in achieving high accuracy in biomedical text mining tasks like named entity recognition from unstructured text~\cite{anantharangachar2013ontology}. 
TMLs are effective for domain-specific fine-tuning in a transfer learning setting and thus become the de-facto standard for representation learning for text mining and information extraction in NLP. 

\begin{figure*}
	\centering
	\includegraphics[width=0.9\textwidth]{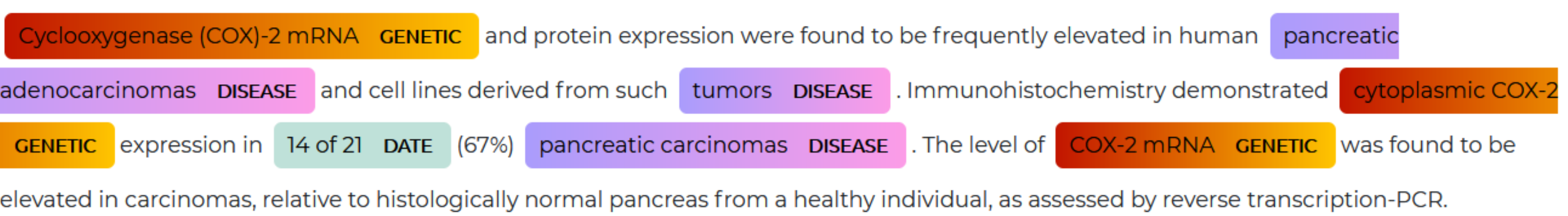}
	\caption{Example of explainable NER, where the model classifies and highlights relevant biomedical entities}
    \label{fig:bio_ner}
\end{figure*}

\begin{figure*}
	\centering
	\includegraphics[width=0.6\textwidth]{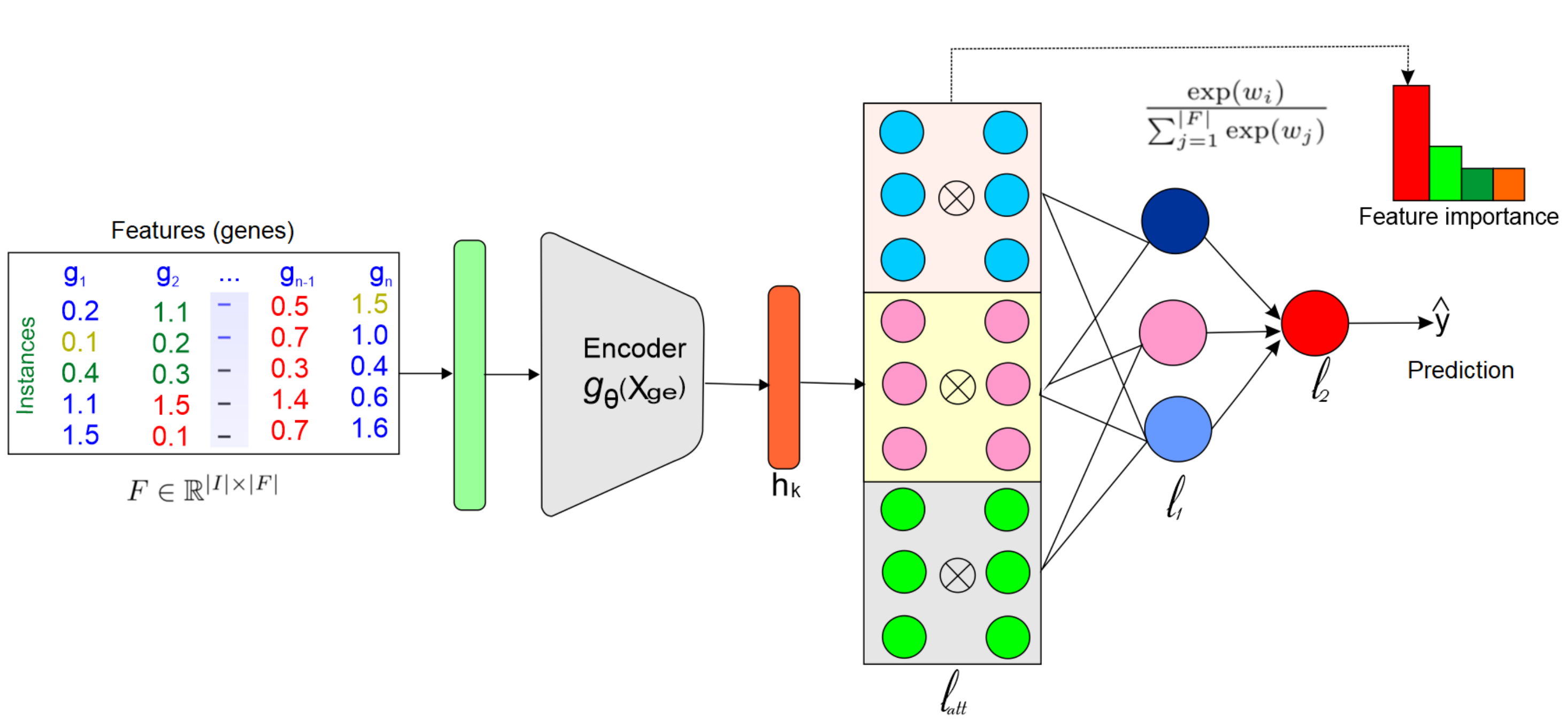}	
	\caption{Example of using SAN for cancer type prediction from gene expression data~(based on~\cite{karim_phd_thesis_2022})}
	\label{fig:k_attention}
\end{figure*}

Besides, the attention mechanism is also increasingly applied. Self-attention network~(SAN)~\cite{vskrlj2020feature} is proposed to identify important features from datasets having a large number of features, indicating that having not enough data can distil the relevant parts of the feature space~\cite{vskrlj2020feature}. TabNet~\cite{arik2021tabnet} is another approach, which uses a sequential attention mechanism to choose a subset of semantically meaningful features to process at each decision step. It visualizes feature importance and how they are combined to quantify individual feature contributions leveraging local and global interpretability. 
Approaches such as SAN and TabNet, place attention layers~(ALs) as hidden layers that map real values to parts of the human-understandable input space~\cite{vskrlj2020feature}:  an element-wise product with $X$ is computed in the forward pass to predict labels $\hat{y}$ in which two consecutive dense layers $l_{1}$ and $l_{2}$ contribute to predictions, $\otimes$ and $\oplus$ are Hadamard product- and summation across $k$ heads. Once the training is finished, weights of ALs are activated using softmax~\cite{vaswani2017attention}. Top-k features then can be extracted as diagonal of $W_{l_{\mathrm{att}}}^{k}$ and ranked w.r.t. their respective weights. 

\subsection{Perturbing black-box model}
Feature-based attributions, game theoretic approach, and sensitivity analysis fall in this category. 

\subsubsection{Feature-based attribution methods} 
~- knowing what features are statistically most important to a model help achieve and generate human-level interpretability. Some features have a higher impact than others. This notion of feature importance~(ref.~\cref{def:feature_importance}) can be computed as permutation feature importance~(PFI). PFI works by randomly permuting a single column in the validation dataset leaving all the other columns intact, where a feature is considered \emph{important} if and only if the model's accuracy drops significantly, thereby increasing the prediction error. A feature is considered \emph{unimportant} if switching its position does not significantly affect the accuracy or performance of the model. 

Feature importance can be conceptualized both locally and globally: effects of a feature for a single prediction, over a large number of samples, or for the overall predictions. Let $x_i$ be a feature in instance $x$. Methods supporting feature importance define an explanation function $g: f \times \mathbb{R}^{d} \mapsto \mathbb{R}^{d}$ and return the importance scores $g(f,x) \in \mathbb{R}^{d}$ for all features for $f$ and a point of interest in $x$. Besides, feature importance can be extracted from tree-based models, e.g., DTs or tree ensembles.  

\begin{figure}
	\centering
	\includegraphics[width=0.45\textwidth]{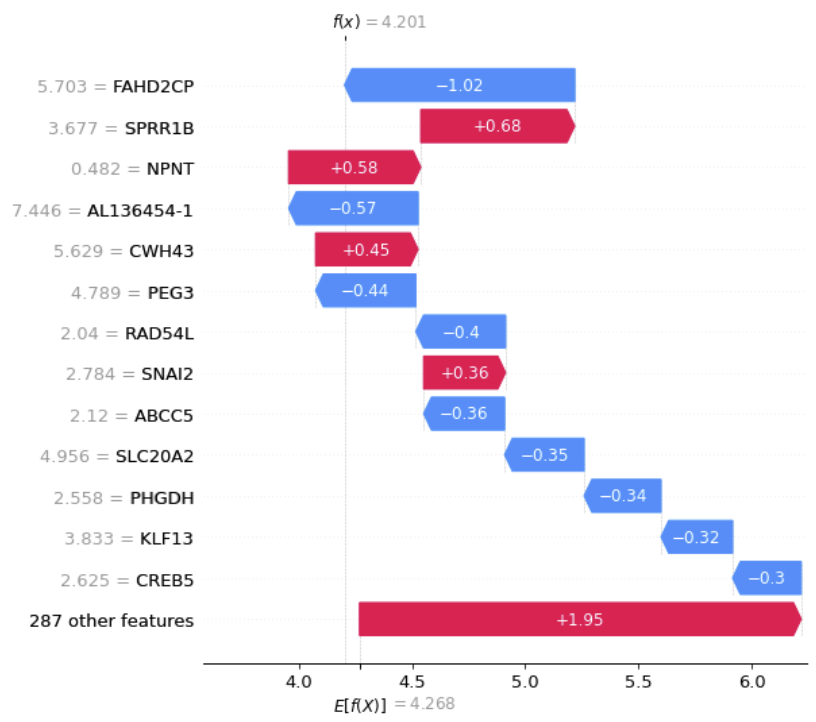}
	\caption{Example of explaining a single cancer type prediction using SHAP waterfall plot: the bottom starts as the expected output value. Each row shows how the positive~(red) or negative~(blue) contribution of individual features pushes the value from expected output over the training dataset to model's output. Positive values imply probabilities of greater than 0.5 that the patient is diagnosed correctly}
    \label{fig:shap_gene_ranking}
\end{figure}


\subsubsection{Game theoretic approach}
~- for a non-linear function~(e.g., a linear model), the order in which features are observed by the model matters. Scott M. \textit{et al.}~\cite{SHAP} showed that tree-based methods are similar to using a single ordering defined by a tree's decision path, yielding inconsistency in PFI~\cite{SHAP}. A widely used perturbation-based method is SHapley Additive exPlanations~(SHAP). SHAP is a game theory-based approach inspired by Shapley values~(SVs). SVs are based on coalitional game theory, i.e., the average marginal contribution of a feature and a way to distribute the gains to its players.~\cite{branzei2008models}. Therefore, SVs give an idea of how to fairly distribute the payout~\cite{molnar2020interpretable}. 

Let $f$ be an interpretable model, $x_i$ be a feature in instance $x$, and $\hat y$ be the prediction. SHAP explains the prediction $\hat y$ by computing the contribution of each feature $x_i$ w.r.t SVs  
and used as measure of feature attributions, where each feature $x_i$ acts as a player. SHAP generate an explanation as~\cite{molnar2020interpretable}:

\begin{equation}
    f\left(z^{\prime}\right)=\phi_{0}+\sum_{i=1}^{C} \phi_{i} z_{i}^{\prime},
\end{equation}

where $z^{\prime} \in\{0,1\}^{C}$ is the coalition vector\footnote{\scriptsize{Coalition vector is a simplified feature representation for tabular data.}} such that the effect of observing or not observing $x_i$ is calculated by setting $z_{i}^{\prime}=1$ or $\left.z_{i}^{\prime}=0\right)$, $\mathrm{C}$ is the maximum coalition size which is equal to the number of input features $M$ and $\phi_{i} \in \mathbb{R}$ is the feature attribution for $x_i$ or SVs. 
The importance $\phi$ of a feature $x_i$ is computed by comparing what model $f$ predicts with and without $x_i$ for all possible combinations of $M-1$ features~(i.e., except for feature $x_i$) in $x$~\cite{lundberg2017consistent}. 

SHAP values explain the output of a function as a sum of the effects $\phi_i$ of each feature being introduced into a conditional expectation. SHAP averages over all possible orderings for computing the mean SVs. If $x_i$ has no or almost zero effect on the predicted value, it is expected to produce an SV of 0. If two features $x_i$ and $x_{i+1}$ contribute equally to the prediction, SVs should be the same~\cite{SHAP}. To compute global importance, absolute SVs per feature across instances are averaged as~\cite{lundberg2017consistent}.

\subsubsection{Sensitivity analysis} 
~- for a \emph{black-box} or interpretable model $f$, sensitivity analysis~(SA) is used to explain a prediction $\hat y$ based on the model's locally evaluated gradient or partial derivatives. Sensitivity $R_{x_i}=\left\|\frac{\partial}{\partial x_{i}} f({x})\right\|$ quantifies the importance of an input feature $x_i$ at a low level~(e.g., image pixel). This measure assumes that the most relevant features are those for which the output is sensitive. Often HMs are plotted to visualize which pixels need to be changed to make the image look similar to the predicted class. However, such an HM does not indicate which pixels are pivotal for a specific prediction, making them not suitable for quantitative evaluation and to validate globally important features. Therefore, SA is more suitable for tabular data to inspect which features a model is more sensitive to. 

To perform SA for a tabular dataset, the value for a feature $x_i$ is changed by keeping other features unchanged. If any change in its value significantly impacts the prediction, the feature is considered to have a high impact and is statistically significant. A new test set $\hat X^{*}$ is often created by applying $w$-perturbation over feature $x_i$, followed by measuring its sensitivity at the global level. To measure the change in the predictions, the mean square error~(MSE) between actual and predicted labels are compared. 
However, SA requires a large number of calculations~(i.e., $N \times M$; $N$ and $M$ are the number of instances and features) across predictions\footnote{\scriptsize{Therefore, some approaches~\cite{karim2022interpreting} recommends making minimal changes to top-k features only, thereby reducing the computational complexity.}}. 

\subsection{Tree, textual, and rules-based approaches} \label{subsec:tree_rules}
\begin{figure*}
	\centering
		\includegraphics[width=0.9\textwidth]{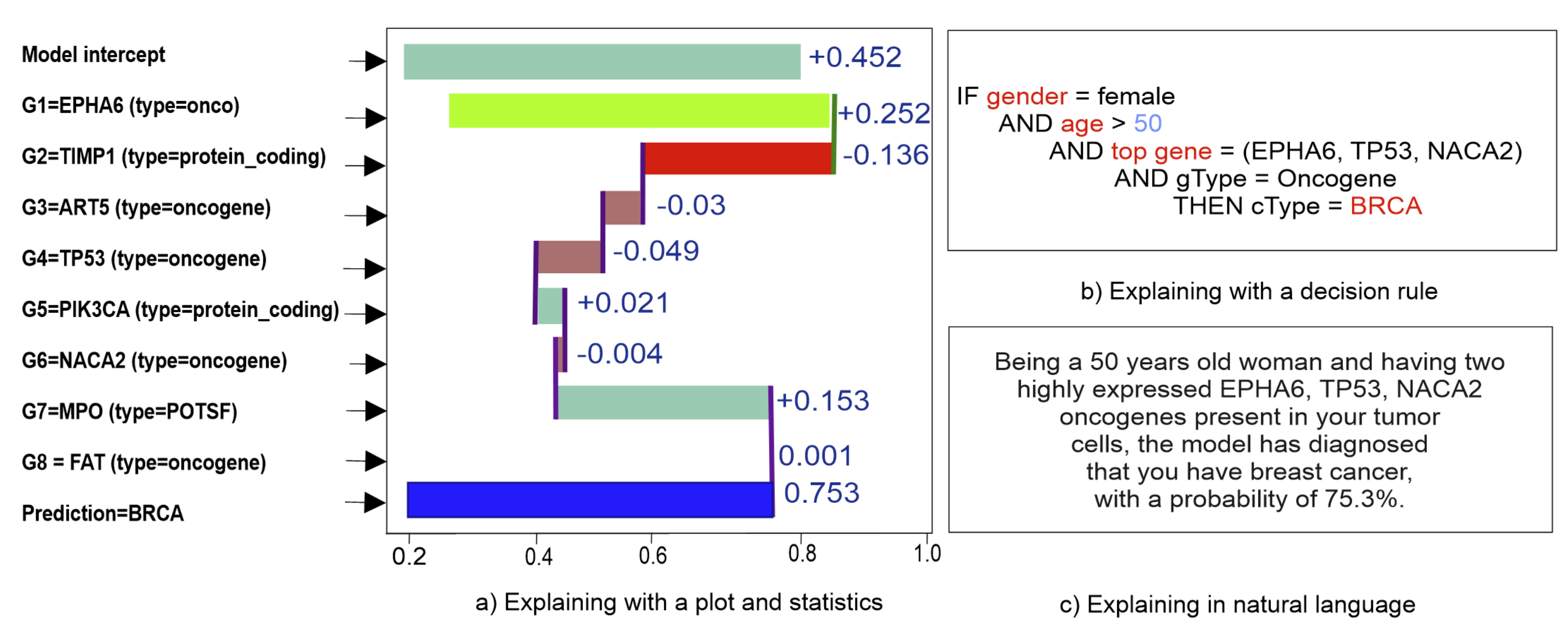}
	    \caption{Explaining a diagnosis decision to a patient is difficult unless it is interpreted simpler way: a) explaining\\ with the plot, b) explaining with a rule, c) explaining in natural human language~\cite{karim_phd_thesis_2022}}
	\label{problem_of_xai:1}
\end{figure*}

For the cancer diagnosis scenario, suppose a DT classifier is used for cancer type prediction. As depicted in \cref{problem_of_xai:1}, consider a test instance where the model predicts that the patient has breast cancer with a probability of 75.3\% based on their GE profile. This prediction shows an average response~(i.e., model intercept) of +0.452. The patient's gender, age, and some marker genes are found to be the most influential features based on their feature impact score. However, combining these concepts to explain the decision may not be easily understandable for all users, such as patients. This is because using plots and charts to explain a decision can be helpful for exploration and discovery, but may be challenging for patients to interpret. Rule-based explanations are more easily understood as they relate the feature values of a sample to its prediction~\cite{ming2018rulematrix}. Using a \emph{decision rule}~(DR) would make it easier to explain the decision in a way that is intuitive to humans. 


DRs have several features: (i) a \emph{general structure} - if certain conditions are met, then make a certain prediction, (ii) the \emph{number of conditions} - there must be at least one feature=value statement in the condition, with no upper limit and additional statements can be added using the ``AND" operator, (iii) \emph{single or multiple DRs} - while multiple rules can be used to make predictions, sometimes a single DR is sufficient to explain the outcome. 
The same decision can be translated into a DR: \emph{``IF gender = female AND age $>$ 50 AND top gene=~(EPHA6, NACA2) AND gType = Oncogene, THEN type = BRCA''}. However, interpreting this for a patient may still be difficult unless they are explained in a human-interpretable way. A simple explanation, in a natural language, could be \emph{``Increased breast cancer is associated with risk for developing lymphedema, musculoskeletal symptoms, and osteoporosis. Being a 50 years old woman and having two highly expressed oncogenes EPHA6 and NACA2 mutated in your tumour cells, you are diagnosed with breast cancer positive, with a probability of 75.3\%.''}. The doctor could further explain why it is the case: \emph{``the model learns attention on the patches from the image and localizes the malignant and normal regions, based on which the diagnosis decision has been made''}.  

\begin{figure*}
	\centering
		\includegraphics[width=0.85\textwidth]{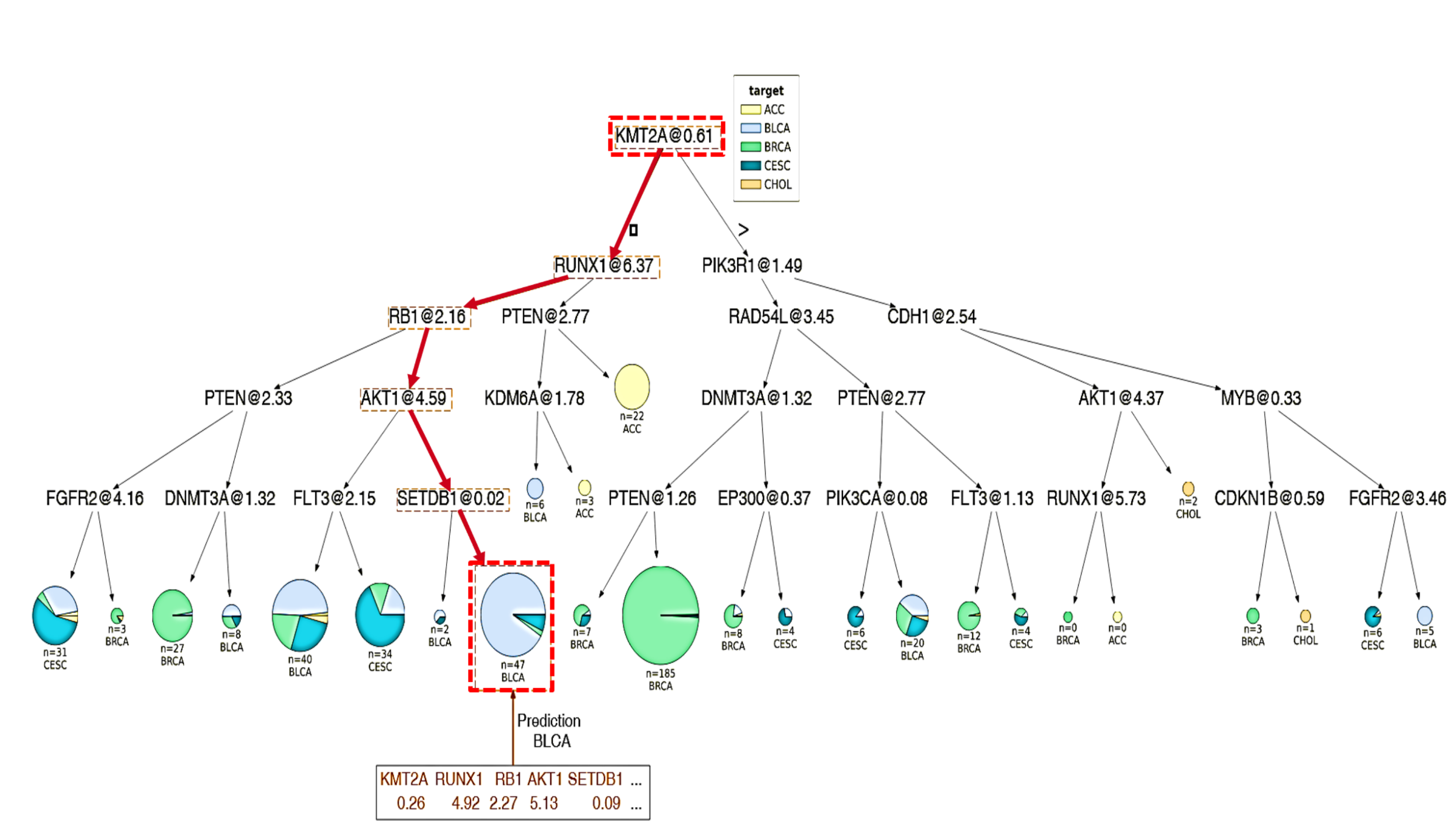}
	    \caption[Decision rules extracted from a root-leaf path]{An example showing how a tree-based model has arrived at decision~(bladder urothelial carcinoma~(BLCA))~\cite{karim_phd_thesis_2022}} 
	    \label{fig:example_DT_path}
\end{figure*}


Local interpretable model-agnostic explanations~(LIME) approximates a \emph{black-box} model via an interpretable model to explain the decisions locally. Anchor~\cite{ribeiro2018anchors} is another rule-based method that extends LIME. It computes DRs by incrementally adding equality conditions on antecedents w.r.t precision threshold~\cite{guidotti2018local}. DRs are arguably the most interpretable predictive models as long as they are derived from intelligible features and the length of the condition is short. However, a critical drawback of rule-based explanations could arise due to overlapping and contradictory rules. Sequential covering and Bayesian rule lists are proposed to deal with these issues. Sequential covering iteratively learns a single rule covering the entire training data rule-by-rule, by removing data points already covered by new rules~\cite{molnar2020interpretable}, while Bayesian rule lists combine pre-mined frequent patterns into a decision list using Bayesian statistics~\cite{molnar2020interpretable}. 

\begin{figure*}
	\centering
		\includegraphics[width=0.9\textwidth]{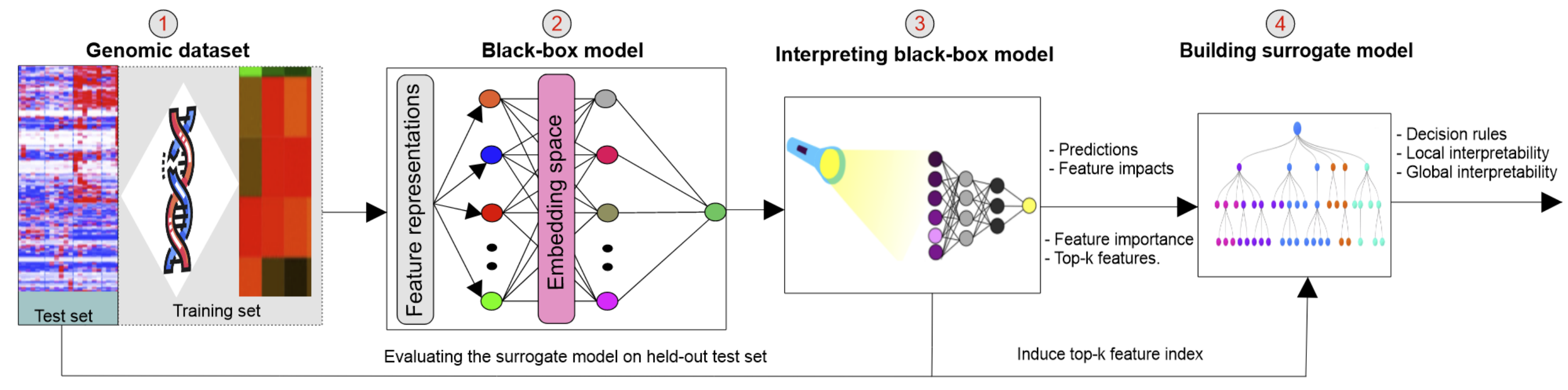}
	    \caption{Creation of interpretable surrogate model based on the multimodal black-box model and using the surrogate model to explain the decision~(grey circles with red numbers represent different steps of the process)~\cite{karim_phd_thesis_2022}}
	    \label{fig:workflow}
\end{figure*}


Tree-based methods such as DTs utilize tree structures, where internal nodes signify feature values relative to Boolean conditions and leaf nodes signify predicted class labels. DTs split the training set into multiple subsets based on the threshold values of features at each node, until each subset contains instances belonging to only one class. Each branch in a DT represents a potential outcome, while the paths from the root to the leaf represent the classification rules. The outcome of a decision tree is the predicted label of a leaf, while the conjunctions of conditions in the \emph{IF} clause match up with different conditions in the path~\cite{guidotti2018survey}. The total of importance values is normalized to 1, with a mean importance of 0 indicating that a feature $x_i$ is highly important, and a mean importance of 0 indicating that the feature is least important. 

As for tree ensembles like boosted trees~(GBTs) or random forests~(RFs), the prediction function $f(\mathrm{x})$ is defined as the sum of individual feature contributions and the average contribution for the initial node in a DT, and the $K$ possible class labels that change along the prediction paths after each split w.r.t information gain~\cite{al2021cdrgi}. DRs then can be derived from a root-leaf path in a DT by starting at the root node of the DT and satisfying the split condition of each decision node, until reaching a leaf node containing the decision.

\subsection{Model surrogation strategies} 
Since interpretability comes at the cost of accuracy vs. complexity trade-off, research has suggested learning a simple interpretable model to imitate a complex model~\cite{molnar2020interpretable}. A \emph{surrogate} or a simple proxy model is often developed to learn a locally faithful approximation of the \emph{black-box}~\cite{stiglic2020interpretability}. Model surrogation is a model interpretation strategy, which involves training an inherently interpretable model by approximating the predictions of a \emph{black-box}~\cite{molnar2020interpretable}.
Depending on the complexity of the problem, the surrogate model is trained on the same data the \emph{black-box} was trained on or on sampled data, as shown in \cref{fig:workflow}. 
Since most important features can be identified by the \emph{black-box} with higher confidence, training an interpretable model on top-k feature space is reasonable. Let $X^{*}$ be a sampled~(e.g., a simplified version of the data containing top-k features only) data for the original dataset $X$ and ${Y}$ be the ground truths. A surrogate model $f$ can be trained on $X^{*}$ and for $Y$. The advantage of model surrogation is that any interpretable model can be used~\cite{molnar2020interpretable}, e.g., LR, DTs or GBT or RF  classifiers\footnote{\scriptsize{Although tree ensembles are complex and known to be \emph{black-boxes}, decision rules can be extracted and FI can be computed from them.}}. 

\subsection{Casual inference and contrastive  explanations}
ML models that produce statistical outputs are based on correlation, not causality~(i.e., focus on association instead of causality). These models map relevant features~(x) to a target variable~(y) based on association, not causality. It is important to note that just because there is a correlation between x and y, it does not mean that x causes y, i.e., \emph{``correlation does not imply causation''}. Recent interpretable methods attempt to address causality by determining which feature caused a specific diagnosis decision made by the model \emph{``Was it a specific feature that caused the diagnosis decision made by the model?''}. Statistical interpretability reveals associations, while causal interpretability answers "what-if" and "why" questions, offering a higher level of interpretability~\cite{moraffah2020causal}. Kim \textit{et al.}~\cite{kim2019learning} suggests learning an oracle model to estimate causal effects for all observed instances and using an interpretable model to approximate the oracle. Another approach is training a black-box to learn causal effects, followed by building an interpretable surrogate model. Lipton \textit{et al.}\cite{lipton2018mythos} argues that a causally interpretable model is often necessary to ensure fairness.

By using a set of DRs users can focus on learned knowledge instead of underlying data representations~\cite{ming2018rulematrix}. Further, humans tend to think in a counterfactual way by asking a question such as \emph{``How would the prediction have been if input $x$ had been different?''}. \emph{Local rule-based explanations}~(LORE)~\cite{guidotti2018local} is an approach that learns an interpretable model by computing a neighbourhood using a genetic algorithm. LORE derives explanations from the interpretable model in the form of DRs and \emph{counterfactuals}~(ref.~\cref{def:Counterfactual_rules}). Partial dependence plot~(PDP) is another way to depict the marginal effect of features on predicted outcomes. PDP allows measuring the change in predictions after making an intervention~(w-perturbations), which can help to discover the features’ causal relationship. 

\subsection{Knowledge-based approaches}
\begin{figure*}[h]
	\centering
		\includegraphics[width=0.9\textwidth]{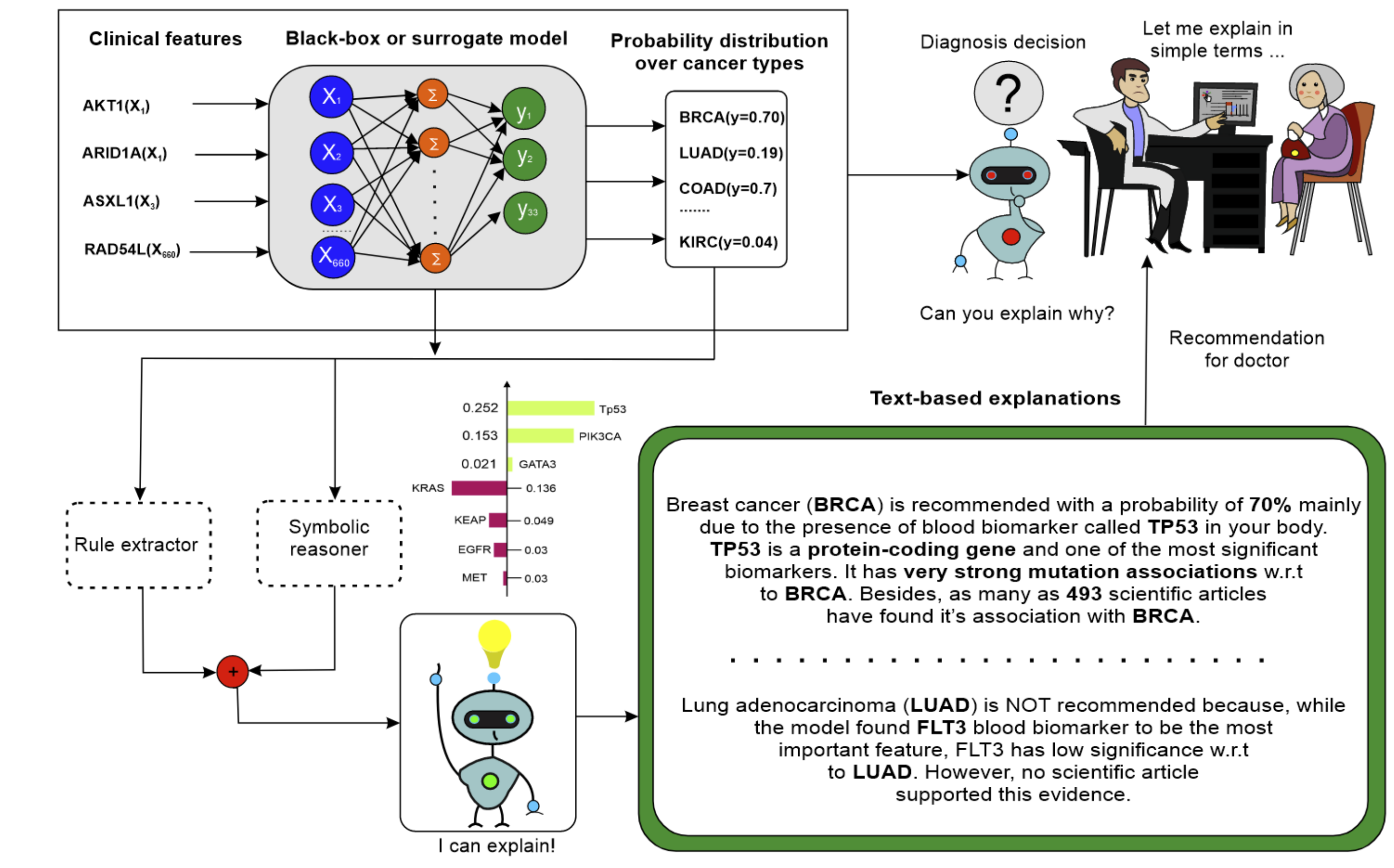}
	\caption{Decision reasoning with rules in textual and natural human language based facts from a domain-knowledge graph~\cite{karim_phd_thesis_2022}}
         \label{fig:decision_reasoning_with_rules}
\end{figure*}

Explanations serve as a bridge between humans and AI systems~\cite{guidotti2018survey}. By allowing for re-enactment and retracing of AI/ML results, interactive ML systems can incorporate human expertise into AI processes, making them more user-friendly~\cite{holzinger2020measuring}. The human-computer interaction~(HCI) community has a long history of advocating for algorithmic transparency in AI systems~\cite{liao2021question}. As shown in \cref{fig:decision_reasoning_with_rules}, an interactive human-AI interface can be created to evaluate the quality of AI explanations. A more concrete example could be developing an explainable chatbot to enable humans to interact with AI systems and receive interactive answers, such as explanations about cancer diagnosis made by ML models.

ML models are typically trained using data and improved through optimization, without incorporating structured domain knowledge. In many scientific research, where a deeper understanding is the goal, relying solely on sophisticated ML methods and high accuracy is not enough. While an ML model may outperform humans in prediction and pattern recognition, it lacks the ability to understand the reasoning behind its decisions. This means that it cannot think like a human and make inferences, abstractions, and connections as a human would~\cite{kapanipathi2020question}. If the model is trained with data that includes expert knowledge or metadata, a well-fitted \emph{black-box} model would have a deeper understanding of the statistically significant features. 


Interpreting a complex \emph{black-box} model can reveal the most important features in a dataset, e.g., in our cancer diagnosis example, the model can outline which oncogenes or protein-coding genes are most significant. However, it cannot determine if these features are biologically significance~(e.g., are all features in the antecedents biologically relevant?). Thus, generated explanations would be based on statistical learning theory, making it difficult to trust the decisions made by the model. Without validation from domain knowledge, these unreliable decisions could have serious consequences in a medical diagnosis, leading to incorrect treatments. To build a reliable and trustworthy AI system, it is essential to understand the biological mechanisms behind carcinogenesis. An interpretable model is not enough and requires rigorous clinical validation with domain expertise before being used in a clinical setting.


The integration of an ML model with a knowledge-based system would provide human operators with reasoning and question-answering capabilities\footnote{\scriptsize{This paradigm, which is being emerged as \emph{neuro-symbolic AI} combines both connectionist AI and symbolic AI paradigms together.}}. Domain experts may need to rely on latest findings from a vast array of sources, e.g., knowledge and facts about drugs, genes, and their mechanism are spread across a huge number of structured~(knowledge bases~(KBs)) and unstructured~(e.g.,  articles)~\cite{karim_phd_thesis_2022,karim2022question}. 
These sources can play a crucial role in understanding biological processes, such as diseases, to inform the development of prevention and treatment strategies~\cite{karim2022question}. The extraction and integration of facts, such as simple statements like \textit{``TP53 is an oncogene"} or quantified statements like \textit{``oncogenes are responsible for cancer"}, into a knowledge graph~(KG) involves three steps: \emph{named entity recognition}~(NER) to identify named entities in biomedical texts, \emph{entity linking} to link the extracted entities with the concepts in a domain-specific KB, and \emph{relation extraction}.

NER is recognizing domain-specific proper nouns in a biomedical corpus. It can be performed by fine-tuning a domain-specific BERT variant such as BioBERT~\cite{BioBERT} and SciBERT~\cite{SciBERT} on relevant articles. For example, for the abstract: \emph{``Cyclooxygenase~(COX)-2 mRNA and protein expression were found to be frequently elevated in human pancreatic adenocarcinomas and cell lines derived from such tumours. Immunohistochemistry demonstrated cytoplasmic COX-2 expression in 14 of 21~(67\%) pancreatic carcinomas. The level of COX-2 mRNA was found to be elevated in carcinomas, relative to the histologically normal pancreas from a healthy individual, as assessed by reverse transcription-PCR.''}, an interpretable BioBERT~\cite{lee2020biobert}-based NER model would be able to recognize named entities classified as diseases, chemical or genetic, as shown in, where named entities are highlighted with different HMs.  
These facts can be extracted as a triplet fact in the form of $(u,e,v)=(\mathit{subject},\mathit{predicate},\mathit{object})$, where each triple forms a connected component of a sentence for the KG, e.g., for the sample text: \emph{``TP53 is responsible for a disease called BRCA\footnote{\scriptsize{Breast cancer}}. TP53 has POTSF\footnote{\scriptsize{Proto-oncogenes with tumour-suppressor function.}} functionality, which is mentioned in numerous PubMed articles''}, \emph{TP53}, \emph{disease}, \emph{BRCA}, \emph{POTSF}, and \emph{PubMed} are the extracted named entities. Then, by linking entities with the concept in a domain-specific ontology, the following relation triples can be generated and integrated into a KG: \texttt{(TP53, causes, BRCA), (TP53, hasType, POTSF), (BRCA, a, Disease), (POTSF, hasEvidence, PubMed)}. 

AI could benefit from external knowledge to support domain experts in understanding why the algorithms came up with certain results~\cite{tiddi2022knowledge}. Tiddi \textit{et al.}~\cite{tiddi2022knowledge} outlined that decisions in the form of counterintuitive predictions could be more understandable if a model provides evidence in the form of explanations based on external, machine-readable knowledge sources, e.g., ontologies can be used to model a component of the AI system to automatically compose explanations exposing different forms of knowledge in order to address a variety of tasks performed by the agent. Therefore, full potential of an AI system can only be exploited by integrating domains- and human knowledge~\cite{tocchetti2022role,zhu2022reasonchainqa,rajabi2022knowledge}. 

A KG can be viewed as the discrete symbolic representations of knowledge~\cite{hogan2020knowledge}. Inference rules~(IRs) are a straightforward way to provide automated access to deductive knowledge~\cite{hogan2020knowledge}. An IR encodes IF-THEN-style consequences: $p \rightarrow y$, where both body and head follow graph patterns in KG.
Therefore, reasoning over a KG using a reasoner would naturally leverage the symbolic technique and would allowing question answering and reasoning\footnote{\scriptsize{See supporting Jupyter notebook in the GitHub repo.}}. It helps deduce implicit knowledge from existing facts~\cite{futia2020integration}, e.g., assuming two facts \emph{``TP53 is an oncogene; oncogenes are responsible for cancer''} are already present, reasoning over the KG would entail an extended knowledge \emph{``TP53 is responsible for cancer''}~\cite{hogan2020knowledge}. 
Then, a doctor with their expertise and by combining the facts from KG, could explain the decision with additional interpretation~(e.g., biomarkers and their relevance w.r.t specific cancer types)~\cite{karim_phd_thesis_2022}. In such a setting, a neuro-symbolic AI system can generate casual reasoning if the KG contains sufficient knowledge~\cite{xian2019reinforcement}. 

\subsection{Measure of interpretability and explainability} \label{sec:xai_measure}
When it comes to providing qualitative or quantitative measures of explainability, there is little consensus on what interpretability in ML is and how to evaluate it for benchmarking~\cite{holzinger2020measuring}. Current interpretability evaluation falls into two categories. The first category evaluates explainability in a quantifiable way: a domain user/expert first claims that some model family, e.g., linear models, rule lists are interpretable and then presents algorithms to optimize within that family~\cite{miller2018explanation}. The second category evaluates the explainability in the context of its applications from a qualitative point of view.  
These approaches rely on the notion of \emph{``you will know it when you see it''}. 
It leaves many questions unanswerable, e.g., \emph{``are all models in all defined-to-be-interpretable''?} or \emph{``are all model classes equally interpretable?''}. However, answers to these questions can only be realized w.r.t some metrics that could qualitative- or quantitatively measure the quality of explanations. 

Kusner \textit{et al.}~\cite{kusner2017counterfactual} proposed a metric for measuring how fair decisions w.r.t counterfactuals. A decision $\hat y$ is fair for instance $x$ if the prediction is the same for both the actual world and the counterfactual world in which the instance belonged to a different demographic group. Since surrogate models are often used to explain the decisions, the quality of the explanations depends on the surrogate's predictive power~\cite{karim2022interpreting}. To measure how well a surrogate $f$ model has replicated a \emph{black-box} $f_b$, R-squared measure~($R^2$) is calculated as an indicator of goodness-of-fit. It is conceptualized as the percentage of variance of $f_b$ captured by the surrogate~\cite{molnar2020interpretable}: 

\begin{itemize}
    \item If $R^2$ is close to $1$~(low error), the surrogate model $f$  approximates the behavior of the \emph{black-box} very well. Hence, $f$ can be used instead of black-box model $f_b$. 
    \item If $R^2$ is close to 0~(high error), the surrogate fails to approximate the \emph{black-box}, hence should not replace $f_b$. 
\end{itemize}

Literature~\cite{yu2019rethinking,deyoung2019eraser} outlined rational explanation as a key to understanding an AI system since translating the rationales results into usable and understandable formats. Deyoung \textit{et al.}~\cite{deyoung2019eraser} proposed comprehensiveness and sufficiency to measure the quality of explanations in NLP. Comprehensiveness signifies whether all features needed to make a prediction, while sufficiency signifies whether extracted rationales contain sufficient signal to support the prediction. These metrics are based on the concept of rationales introduced by Zaidan \textit{et al.}~\cite{zaidan2007using} in NLP, where a human annotator would highlight a part of the texts that could support the labelling decision. 

Let $f\left(x_{i}\right)_{c}$ be the original prediction probability for a model $f$ for the class $c$. To measure the sufficiency, Deyoung et al.~\cite{deyoung2019eraser} proposed to create contrasting example $\tilde{x}_{i}$ for each sample $x_{i}$, by removing predicted rationales $r_{i}$: 

\begin{align}
    s  =  f\left(x_{i}\right)_{c} f\left(r_{i}\right)_{c},
    \label{eq:sufficiency}
\end{align}

where $s$ measures the degree to which extracted rationales are adequate for the model $f$ for making the prediction~\cite{deyoung2019eraser}. Let $f\left(x_{i} \backslash r_{i}\right)_{c}$ be the predicted probability of $\tilde{x}_{i}\left(=x_{i} \backslash r_{i}\right)$. Thus, it is expected the prediction to be lower on removing the rationales, where comprehensiveness $e$ is calculated as~\cite{deyoung2019eraser}: 

\begin{align}
    e  = f\left(x_{i}\right)_{c}-f\left(x_{i} \backslash r_{i}\right)_{c}.
\end{align}

The same idea can be conceptualized for our cancer example w.r.t leave-one-feature-out: the rationale can be computed based on the number of extracted features~(e.g., top-k genes) divided by the number of features in $x$. A prediction $\hat y$ is a match if the overlap with any of the ground truth rationales $r_i \geq \sigma$, where $\sigma$ is a predefined threshold set by a domain experts. A high value of comprehensiveness implies that the rationales were influential w.r.t prediction $\hat y$. If an AI system is useful, then it must somehow be interpretable and able to provide human-understandable explanations~\cite{bhatt2020explainable}. 

\begin{table*}
    \centering
    \captionsetup{justification=centering}
    \caption{SCT and it's usability questionnaires with interpretation~(adopted from~\cite{holzinger2020measuring})}
    \label{table:usability_questionnaires}
    \scriptsize{
    \begin{tabular}{l|l}
        \hline
         \textbf{Usability question}  & \textbf{Interpretation}  \\
        \hline
        \textbf{Factors in data}  & Data included all relevant known causal factors with sufficient precision and granularity \\ \hline
        \textbf{Understood} & Explanations are within the context of my work \\  \hline
        \textbf{Change detail level} & Level of detail can be changed on demand \\  \hline
        \textbf{Understanding causality} & No support was necessary to understand the explanations \\  \hline
        \textbf{Use with knowledge} & Explanations helped me understand the causality \\  \hline
        \textbf{No inconsistencies} & Explanations could be used with my knowledge base \\  \hline
        \textbf{Learn to understand} & No inconsistencies between explanations \\  \hline
        \textbf{Needs references} & Additional references were not necessary in the explanations \\  \hline
        \textbf{Efficient} & Explanations were generated in a timely and efficient manner. \\
        \hline
    \end{tabular}}
\end{table*}



System casuability scale~(SCS)~\cite{holzinger2020measuring} is another measure based on the notion of causability~\cite{holzinger2019causability}. It is proposed to quickly determine whether and to what extent an explainable user interface, an explanation, or an explanation process itself is suitable for the intended purpose~\cite{holzinger2020measuring}. 
SCS combines with concepts adapted from a widely accepted usability scale and SCS is computed based on responses from ten usability questionnaires listed in \cref{table:usability_questionnaires}. SCS system uses a 5 point scale: rating 1 = strongly disagree; 2 = disagree; 3 = neutral; 4 = agree; 5 = strongly agree. SCS is measured by dividing the acquired ratings by the total ratings, i.e., $SCS  = \sum_{i=1}^{10} \text {Rating}_{i} / 50$. 

\section{Interpretable ML Tools and Libraries} \label{sec:xai_library}
Following the timeline of interpretable and XAI methods, numerous libraries and tools have been developed. Majority of these tools are developed for general purpose problems. Besides, some tools were customized to solve domain-specific problems. 

\subsection{General purpose XAI tools}
Most of these tools and libraries are developed with a view to improving the interpretability and explainability of \emph{black-box} ML models, covering general-purpose problems in computer vision, text mining, or structured data, and are based on well-known interpretable ML methods such as LIME~\cite{LIME}, model understanding through subspace explanations~(MUSE)~\cite{lakkaraju2019faithful}, SHAP~\cite{SHAP}~(and its variants such as kernel SHAP and tree SHAP), partial dependence plot~(PDP), individual conditional expectation~(ICE), permutation feature importance~(PFI), and counterfactual explanations~(CE)~\cite{wachter2017counterfactual}. Following are some widely used and general purpose interpretable tools and libraries, and links and their GitHub repositories: 

\begin{itemize}
    \scriptsize{
        \item \textbf{DeepLIFT}: \url{https://github.com/kundajelab/deeplift} 
        \item \textbf{Xplique}: \url{https://github.com/deel-ai/xplique/}
        \item \textbf{DALEX}: \url{https://dalex.drwhy.ai/python/}
        \item \textbf{Alibi}: \url{https://github.com/SeldonIO/alibi}
        \item \textbf{SHAP}: \url{https://github.com/slundberg/shap}
        \item \textbf{LIME}: \url{https://github.com/marcotcr/lime}
        \item \textbf{PyTorch-Grad-CAM}: \url{https://github.com/jacobgil/pytorch-grad-cam}
        \item \textbf{ELI5}: \url{https://github.com/TeamHG-Memex/eli5}
        \item \textbf{InterpretML}: \url{https://github.com/interpretml/interpret}
        \item \textbf{Interpret-Text}: \url{https://github.com/interpretml/interpret-text}
        \item \textbf{ExplainerDash}: \url{https://github.com/oegedijk/explainerdashboard}
        \item \textbf{CNN viz}: \url{https://github.com/utkuozbulak/pytorch-cnn-visualizations} 
        \item \textbf{iNNvestigate}: \url{https://github.com/albermax/innvestigate} 
        \item \textbf{DeepExplain}: \url{https://github.com/marcoancona/DeepExplain} 
        \item \textbf{Lucid}: \url{https://github.com/tensorflow/lucid} 
        \item \textbf{TorchRay}: \url{https://facebookresearch.github.io/TorchRay/} 
        \item \textbf{Captum}: \url{https://captum.ai/} 
        \item \textbf{AIX360}: \url{https://github.com/Trusted-AI/AIX360}
        \item \textbf{BERTViz}: \url{https://github.com/jessevig/bertviz}.
    }
\end{itemize}

\subsection{Customizing XAI Methods for Bioinformatics} \label{sec:customize_xai}
 General purpose XAI tools are not specialized to tackle bioinformatics problems by default. This makes the direct application to bioinformatics problems challenging without customization and domain-specific adaptation. For instance, LIME is mostly suitable for tabular data, even though it supports image, text and tabular data. Now in order to perform time series classification, LIME needs to be extended so that it is able to deal with time series data. One approach could be perturbing~(e.g., parts of its features are \emph{switched off}, pixels greyed out) input instance several times before feeding into the \emph{black-box}. The approximating model then learns which features to have the most influence on the final prediction.


Methods like SHAP were developed to determine how each input alters the model prediction. However, interpretation of models trained from biological sequences remains more challenging because model interpretation often ignores ordering of inputs. ``Positional SHAP"~(PoSHAP)~\cite{PoSHAP} is proposed to interpret models trained from biological sequences by exploiting SHAP to generate positional model interpretations. Study has found that PoSHAP helps improve interpretability for DNN models trained on biological sequences in a variety of tasks such as peptide binding motifs, reflected known properties of peptide CCS, and provided new insights into interpositional dependencies of amino acid interactions. A ML model can be trained on biological sequence data as an input to predict peptide collisional cross section and to predict peptide binding affinity to major histocompatibility complex~(MHC) isoforms. However, in order to enable positional interpretation for the predictions, input indexes need to be added to the inputs to calculate the SVs from the models. Then, for every sequence, top-5 matches can be identified to a given position weight matrix~(PWM) and also investigate the total importance assigned to the positions underlying those matches. 


Studies have found that domain shift in imaging can result in two major differences in image quality and appearance, with sharpening, changes in contrast, brightness, and intensity~\cite{zhang2020generalizing}. 
Therefore, in case of bioimaging~(e.g., MRIs, CT, and X-rays), modality-specific preprocessing steps~(e.g., rescaling and horizontal flipping in case of radiographs, while MRIs may require contrast enhancement, intensity regulation, noise elimination) are necessary. Further, the network weights should not be initialized with ImageNet like pretrained models, as they often contain photos of general objects, otherwise it would activate the internal representation of the network's hidden layers with geometrical forms, colorful patterns, or irrelevant shapes that are usually not present in radiography images~\cite{karim2021deepkneeexplainer}. These are necessary to improve model generalization capability as well as not to influence imaging interpretability methods such as Grad-CAM++ and LRP. The histocartography~\cite{jaume2021histocartography} is library designed to facilitate the development of graph-based computational pathology pipelines, where Grad-CAM++ were extended as GraphGradCAM and GraphGradCAM++ for cell graph explainer to generate an explanation to highlight salient nodes. 


TLMs such as BERT were originally pre-trained on English corpus such as Wiki dump, newspapers, and books. Thus, without domain-specific finetuning they cannot be directly applied to biomedical texts containing a considerable number of domain-specific proper nouns~\cite{karim_phd_thesis_2022}. BioBERT~\cite{BioBERT} or SciBERT~\cite{SciBERT} can be initialized with a case-sensitive version of BERT, followed by finetuning their weights on PubMed abstracts to perform the NER and enity linking tasks. Local explanations for individual predictions then can be provided by highlighting important features in an input biomedical text sample in a post-hoc fashion. For instance, the relevance score as a measure of importance can be computed with relevance conservation LRP~\cite{arras2017explaining}. The output value for each bio-entity predicted then can be back-propagated layer-wise onto the token level, where token relevances can be visualized with HMs~(e.g., \cite{karim2020deephateexplainer}) or feature attributions~(BERT-LRP~\cite{wu2021explaining}). 

Sometimes trained models may need to be converted to support the above general purpose XAI tools. For instance, in order to embed the capability for computing importance scores using DeepLIFT, a Keras model need to convert as deeplift model. Further, not each form of explanations can be generated using a single tool, which implies that a single tool or even multiple may need to be customized or be combined, to develop XAI applications for bioinformatics.

Following are examples of customized XAI tools that can be used in the contexts of biomedical/bioinformatics use cases: 

\begin{itemize}
    \scriptsize{
        \item \textbf{Bio-NER}: \url{https://librairy.github.io/bio-ner/}
        \item \textbf{LIME for time}: \url{https://github.com/emanuel-metzenthin/Lime-For-Time}
        \item \textbf{Positional-SHAP}: \url{https://github.com/jessegmeyerlab/positional-SHAP}
        \item \textbf{BERT-LRP}: \url{https://github.com/frankaging/BERT-LRP} 
        \item \textbf{Histocartography}: \url{https://github.com/BiomedSciAI/histocartography} 
    }
\end{itemize}

\section{Conclusions}\label{sec:con}
Interpretability is a key to generate insights on why and how certain predictions are made by a model. 
In this paper, we discussed importance of interpretability in bioinformatics and provided a comprehensive overview of interpretable methods. 
Via several examples of bioimaging, genomics, and biomedical texts, we demonstrated how bioinformatics research could be benefited from interpretability and different means of explanation types~(e.g., rules, plots, heatmaps, textual, and knowledge-based). Besides, we analyzed existing interpretable ML tools and libraries that can be employed to improve interpretability for complex bioinformatics research problems. 


Although interpretability could contribute to transparent AI~\cite{kim2019learning}, interpretability alone cannot guarantee the trustworthiness of an AI system. The benefits of explainability still need to be proven in practical settings. The EU's guidelines on AI robustness and explainability~\cite{hamon2020robustness} emphasize three key elements for the proper utilization of AI: transparency, reliability, and safeguarding of individual data. We argue that there are other important considerations too in the development and deployment phases of an AI system -- especially for mission-critical applications like clinical use~\cite{kazim2020explaining}: 

\begin{enumerate}
    \item Before building an interpretable model, decision-makers and experts should consider factors such as: i) what kind of data to use - imaging, text, tabular, graphs?, ii) what types of explanations to provide, e.g., visual-, tree-, rule-, or textual?, iii) how could full potentials of global interpretability be achieved to tailor or generalize the model better for unseen data?, iv) how could local interpretability be used if the model fails, so it can be diagnosed before re-training?, and v) what potential impacts the model could have on the targeted users?~\cite{kazim2020explaining}. 
    
    \item One of the first steps to improving an AI system is to understand its weaknesses: the better we understand which factors cause it to make right predictions or fail, the easier it becomes to improve it~\cite{miller2018explanation}. However, such weakness analysis on \emph{black-box} or interpretable models is not straightforward~\cite{bhatt2020explainable}, but requires close monitoring and debugging by zooming individual data points. 
    
    \item Explanations generated by AI-assisted decisions may not only reveal commercially sensitive information, but also the inner workings of an ML model~\cite{kazim2020explaining}. There are potential dangers associated with the rationale, fairness, and types of data explanation, such as information on how similar individuals were treated and details on the input data used to make a decision. To mitigate these risks, it is crucial to restrict the amount of detail provided, such as feature weightings or importance, and to thoroughly evaluate the risk as part of a data protection.

    \item AI systems may be vulnerable to adversarial attacks, bias against underrepresented groups, and inadequate protection of individual data, which not only negatively affects user experience but also undermines societal trust. To address these issues, an AI system must be robust to potential adversaries by taking both reactive and proactive measures. It is crucial to ensure that the predictions remain consistent and reliable even in the presence of minor variations in input data, e.g., adding small amounts of noise to input should not drastically change the predictions.
    
    \item An AI system should not only provide \emph{meaningful information}~(e.g., clinical outcomes) and clarify the reasoning behind its decisions through supporting explanations, but it should also offer \emph{insights} like treatment recommendations. Further, biological relevance of important factors needs to be validated both clinically and based on domain knowledge, e.g., oncologists can combine their expertise with evidence from a domain knowledge graph.     
\end{enumerate}

Finally, we believe that the techniques and methods discussed in the paper offer valuable insights into interpretable machine learning and explainable AI. We hope that it will benefit domain experts such as doctors and data scientists, lay people like patients, and stakeholders, ultimately accelerating bioinformatics research.

\subsection{{\textbf{Key points}}}

\begin{itemize}
    \item \textbf{Ante-hoc or post-hoc interpretable methods?} Interpretable models, such as linear models, should be the first choice for building and explaining due to their simplicity and ease of comprehension. However, in complex modeling scenarios, a surrogate model may not accurately reflect the behavior of a \emph{black-box} model. This could result in incorrect decisions if the surrogation process is performed without proper evaluation~\cite{karim2022interpreting}. It is therefore recommended to build a \emph{black-box} model first, then incorporate interpretable ML logic and use the resulting interpretable model to explain decisions. The latter approach can combine both ante-hoc and post-hoc approaches in a single pipeline.
    
    \item \textbf{Local or global explainability?} 
    Global explainability is important for monitoring and having a holistic view of a model's performance and interpretability~(e.g., what features across training instances are more important to the model?). Local explainability is important for individual instances without providing a general understanding of the model. Further, local explainability helps in diagnosing which factors contributed in wrong predictions. 

    \item \textbf{What ML models and explanations types?} Although the choice depends on problems, requirements, data types, as well as target users~(decision recipients), e.g., decision rules and counterfactuals are more effective to provide intuitive explanations compared to visualization explanations~(the former is suitable for tabular data, the latter is for imaging)~\cite{molnar2020interpretable}, an AI system needs to build such that relevant information can be extracted for a range of explanation types~\cite{karim2022interpreting}. Via a user-study, Tiddi \textit{et al.}~\cite{tiddi2022knowledge} identified that different types of explanations required at the different steps of the automated reasoning for scenarios like cancer diagnosis and treatment in clinical settings: ``everyday explanations" for diagnosis, ``trace-based explanations" for planning the treatment, ``evidence-based explanations" to provide scientific evidence from existing studies, and ``counterfactual explanations" to allow clinicians to add/edit information to view a change in the diagnosis or treatment recommendation. In other words, since one-size-may-not-fit-all, several explanation types out of plot-, tree-, rule-, evidence-, or text-based may need to combine.  
    
    \item \textbf{Accuracy or explainability?} Balancing accuracy and explainability can be challenging, given the varying level of complexity of the problem, types of data, and amount of data. In critical scenarios like healthcare, both accuracy and reliability are key considerations, as mediocre performance is not acceptable. However, when a complex problem needs to be solved and an efficient model needs to be built, attaining higher accuracy may not be the highest priority, as long as the solution is still understandable.  

    \item \textbf{Can explainability help mitigate unfairness in the decision-making process?} Interpretability and explainability are crucial factors in identifying sensitive features that may lead to discriminatory outcomes in decision-making. They allow for proactive measures to be taken to prevent unfair treatment of certain groups or populations based on such attributes. 
    
    \item \textbf{Is explainability always necessary?} Increased interpretability leads to better understanding and simplifies the task of explaining predictions to end users~\cite {stiglic2020interpretability}. However, not all predictions need to be explainable. In some application domains, such as disease diagnosis in translational bioinformatics, addressing AI learning security or fixing learning flaws may be more important than creating an interpretable model~\cite{han2022challenges}.
    
    \item \textbf{Human-in-the-loop and domain knowledge?} HCI researchers should work closely with domain experts such as data scientists and doctors throughout the development phase, utilizing only cutting-edge ML algorithms and interpretable methods. This collaboration will enable the AI developers to design user interfaces that allow for human operators from various domains to ask ``why", ``how", and ``what-if" questions and receive clear and contrasting explanations in diverse formats.
\end{itemize}

\section*{{\textbf{Conflict of Interest}}}
No conflict of interest to declare for this article.

\section*{{\textbf{Acknowledgment}}}
This article is loosely based on Ph.D. dissertation titled \emph{``Interpreting black-box machine learning models with decision rules and knowledge graph reasoning''}~\cite{karim_phd_thesis_2022} by the first author. 

\section*{\textbf{Abbreviations and Acronyms}}

{\small{
\begin{itemize}
    \item Attention layers~(ALs) 
    \item Artificial intelligence~(AI) 
    \item Autoencoders~(AEs)
    \item Bayesian rule lists~(BRL)
    \item Bayesian case Model~(BCM)
    \item Convolutional neural network~(CNN)
    \item Concept activation vectors~(CAVs)
    \item Counterfactual explanations~(CE)
    \item Contrastive layer-wise relevance propagation~(CLRP)
    \item Convolutional autoencoder~(CAE)
    \item Class activation mapping~(CAM)
    \item Causal concept effect(CaCE)
    \item Casual inferencing~(CI)
    \item Deep learning~(DL)
    \item Deep neural networks~(DNNs)
    \item Decision rules~(DRs)
    \item Decision tree~(DT)
    \item Explainable artificial intelligence~(XAI)
    \item Feature maps~(FMs)
    \item General data protection regulation~(GDPR)
    \item Generative discriminative models~(GDM)
    \item Global attribution mapping~(GAM)
    \item Gene expression~(GE)
    \item Guided back-propagation~(GB)
    \item Gradient-weighted class activation mapping~(Grad-CAM++)
    \item Global average pooling~(GAP)
    \item Globally averaged gradients~(GAG)
    \item Gradient boosted trees~(GBT)
    \item Graph neural networks~(GNNs)
    \item Heat maps~(HMs)
    \item Isometric mapping~(Isomap)
    \item Individual conditional expectation~(ICE)
    \item K-nearest neighbour~(KNN)
    \item Layer-wise relevance propagation~(LRP)
    \item Local interpretable model-agnostic explanations~(LIME)
    \item Local rule-based explanations~(LORE) 
    \item Long short-term memory~(LSTM)
    \item Machine learning~(ML)
    \item Model-agnostic counterfactual explanation~(MACE)
    \item Model understanding through subspace explanations~(MUSE)
    \item Natural language processing~(NLP)
    \item Neural additive models~(NAMs)
    \item Principal component analysis~(PCA)
    \item Partial dependence plots~(PDP)
    \item Peak response mapping~(PRM)
    \item Propositional self-attention networks~(PSAN)
    \item Prediction difference analysis~(PDA)
    \item Permutation feature importance~(PFI)
    \item Random forest~(RF)
    \item Rectified linear unit~(ReLU)
    \item Saliency maps~(SM)
    \item Self-attention network~(SAN)
    \item System causability scale~(SCS)
    \item Sequential covering~(SC)
    \item SHapley Additive exPlanations~(SHAP)
    \item Sensitivity analysis~(SA)
    \item Salient relevance~(SR)
    \item Spectral relevance analysis~(SpRAy)
    \item Shapely values~(SVs)
    \item Singular value decomposition~(SVD).
    \end{itemize}
    }
}

\section*{Author details}
\noindent \textbf{Md. Rezaul Karim} is a Senior Data Scientist at Fraunhofer FIT and a Postdoctoral Researcher at RWTH Aachen University, Germany. Previously, he worked as a Machine Learning Researcher at Insight Centre for Data Analytics, University of Galway, Ireland. Before that, he worked as a Lead Software Engineer at Samsung Electronics, South Korea. He received his PhD from RWTH Aachen University, Germany; M.Eng. degree from Kyung Hee University, South Korea; and B.Sc. degree from the University of Dhaka, Bangladesh. His research interests include machine learning, knowledge graphs, and explainable AI~(XAI) with a focus on bioinformatics. \\

\noindent \textbf{Tanhim Islam} is a Data Scientist at CONET Solutions GmbH, Germany. He received his BSc. degree in Computer Science from the BRAC University, Bangladesh and MSc. from RWTH Aachen University, Germany. His research interests include applied machine learning, NLP, and explainable AI. \\

\noindent \textbf{Oya Beyan} is a Professor of medical informatics at University of Cologne, Faculty of Medicine, and University Hospital Cologne. Prof. Beyan is the director of Institute for Biomedical Informatics which aims to support data-driven medicine and digital transformation in healthcare. Prof. Beyan's research team focuses on health data reusability, semantic interoperability, and data science towards continuous improvement of healthcare through innovation and creating new knowledge. Prof. Beyan co-leads the MEDIC data integration center at Medical Faculty of Cologne, where multi-modal health data is integrated and reused for research. She is also affiliated with Fraunhofer FIT, where she leads a research and innovation group for FAIR data and distributed analytics. \\

\noindent \textbf{Christoph Lange} is the Head of Data Science and Artificial Intelligence department at Fraunhofer FIT, Germany, and a Senior Researcher at RWTH Aachen University. He received his Ph.D. in computer science from Jacobs University Bremen, Germany. His research is broadly concerned with knowledge engineering and data infrastructures: choosing the right formalism domain to represent domain knowledge in an expressive and scalable way following Linked Data and FAIR principles, thus enabling and facilitating machine support with sharing, publishing, and collaborative authoring. \\

\noindent \textbf{Michael Cochez} is an Assistant Professor at Vrije Universiteit Amsterdam, the Netherlands. He was formerly a Senior Researcher at Fraunhofer FIT, Germany. He received his Ph.D. and M.Sc. degrees in Mathematical Information Technology from the University of Jyv\"askyl\"a, Finland, and B.Sc. degree in Information Technology from the University of Antwerp, Belgium. His research interests include knowledge representation, deep learning, machine learning, and question answering over knowledge graphs. \\

\noindent \textbf{Dietrich Rebholz-Schuhmann} is a Professor of Medicine at the University of Cologne and the Scientific Director of the Information Center for Life Sciences, German National Library of Medicine~(ZB MED), Germany. He was formerly the Director of the Insight Centre for Data Analytics, and Professor of Informatics at the University of Galway, Ireland. Before that, he was the Director of Healthcare IT at a Heidelberg, Germany-based bioscience company and Group Leader of Semantic Data Analytics at European Bioinformatics Institute~(EMBL-EBI). His research interests include data science, semantics-driven data analytics, biomedical text mining, and bioinformatics. \\

\noindent \textbf{Stefan Decker} is the Chair and Professor for Information Systems and Databases at RWTH Aachen University and Managing Director of Fraunhofer FIT, Germany. He was formerly the Director of the Insight Centre for Data Analytics, and Professor of Informatics at University of Galway, Ireland. Even before that, he worked as Assistant Professor at the University of Southern California and as a Postdoctoral Researcher at Stanford University, USA. His research interests include Semantic Web and linked data, and knowledge representation. 

\bibliographystyle{natbib}
\bibliography{references}

\end{document}